\documentclass[amssymb,12pt,longbibliography,nofootinbib]{revtex4-1}

\def\journal{Phys.~Rev.~A {\bf 108}, 042210 (2023)}
\def\link{https://doi.org/10.1103/PhysRevA.108.042210}

\usepackage{graphicx}
\usepackage{appendix} 
\usepackage{hyperref}
\usepackage{bm,amsmath}
\usepackage{amsthm}
\usepackage{setspace}
\usepackage{slashed,mathrsfs}

\def\BLO{ B ({\cal H}, {\mathbb C}) }

\def\DIM{ N }

\def\I{ \mathrm{i} }

\newcommand{\BRA}[1] { \left\langle #1 \right|}
\newcommand{\CMATRIX}[1] {{\mathbb C}^{ #1 \! \times \! #1}}

\newcommand{\KET}[1] {\left| #1 \right\rangle}
\newcommand{\KETBRA}[1] {\left| #1 \right\rangle \! \left\langle  #1 \right|  }

\newtheorem{lem}{Lemma}

\newtheorem{thm}{Theorem}

\newcommand{\WFRAME}[1] { {\mathbb W}^{ (#1) } }
\newcommand{\SFRAME}[1] { {\mathbb S}_{ #1 } }

\begin{document}
\rightline{\href{\link}{\tt \journal}}
\vspace{0.2in}

\title{The universe as a nonlinear quantum simulation: Large $n$ limit of the central spin model}
\author{\textsc{Michael R. Geller}}
\affiliation{Center for Simulational Physics, University of Georgia, Athens, Georgia 30602, USA}
\date{October 19, 2023}
 
\begin{abstract}
\vskip 0.5in
\centerline{\bf Abstract}
\vskip 0.1in
\begin{spacing}{0.9}
We investigate models of nonlinear qubit evolution based on mappings to an $n$-qubit central spin model (CSM) in the large $n$ limit, where mean field theory is exact. Extending a theorem of Erd\"os and Schlein, we establish that the CSM is rigorously dual to a nonlinear qubit when $n \rightarrow \infty$. The duality supports a type of nonlinear quantum computation in systems, such as a condensate, where a large number of ancilla couple symmetrically to a ``central'' qubit. It also enables a gate-model implementation of nonlinear quantum simulation with a rigorous error bound. 
Two variants of the model, with and without coupling between ancilla, map to effective models with different nonlinearity and symmetry. Without coupling the CSM simulates initial-condition nonlinearity, where the Hamiltonian is a linear combination of 
${\rm tr}(\rho_0 \sigma^x) \sigma^x$, 
${\rm tr}(\rho_0 \sigma^y) \sigma^y$, and
${\rm tr}(\rho_0 \sigma^z) \sigma^z$, 
where 
 $\sigma^x, \sigma^y, \sigma^z$ are Pauli matrices and $\rho_0$ is the initial density matrix. 
With symmetric ancilla coupling it simulates linear combinations of 
${\rm tr}(\rho \sigma^x) \sigma^x$, 
${\rm tr}(\rho \sigma^y) \sigma^y$, and
${\rm tr}(\rho \sigma^z) \sigma^z$, where
$\rho$ is the current state.
This case can simulate qubit torsion, which has been shown by Abrams and Lloyd to enable an exponential speedup for state discrimination in an idealized setting. 
The duality discussed here might also be interesting from a quantum foundations perspective. There has long been interest in whether quantum mechanics might possess some type of small, unobserved nonlinearity. If not, what is the principle prohibiting it? The duality implies that there is not a sharp distinction between universes evolving according to linear and nonlinear quantum mechanics: A one-qubit ``universe" prepared in a pure state $\KET{\varphi}$ at the time of the big bang and symmetrically coupled to ancilla prepared in the same state, would appear to evolve nonlinearly for any finite time $t>0$ as long as there are exponentially many ancilla $n \gg {\rm exp}(O(t))$. 
\end{spacing}
\end{abstract}
\maketitle

There is a growing interest in exploring, as a purely theoretical question, the computational power of hypothetical forms of quantum nonlinearity \cite{MielnikJMP80,PhysRevLett.81.3992,BechmannPLA98,0502072,150706334,211105977,9802051,9803019,0309189,BrunPRL09,BennettPRL09,13033537,13030371,13107301,200907800,220613362,XuPRR22}.
One motivation is the intriguing 1998 paper by Abrams and Lloyd \cite{PhysRevLett.81.3992} arguing that evolution by certain nonlinear Schr\"odinger equations, in an idealized setting, would allow NP-complete problems to be solved efficiently. Meanwhile, there is a growing body of algorithms developed to {\it simulate} nonlinear problems, such as dissipative fluid flow, with a linear quantum computer 
\cite{YukawaAndMiyataPRA13,LeeAndKurzynskiPRA15,JosephPRR20,GaitanNPJQI20,201103185,201106571,201206681,210303804,210507317,DodinPhysPlas21,210710764,210908470,211107486,211200602,211212307,220202188}.  Such algorithms provide a link between linear and nonlinear representations of the same problem, and might teach us something about quantum nonlinearity itself. Here we explore this question in the context of a recent algorithm proposal by Lloyd {\it et al.}~\cite{201106571} for the quantum simulation of nonlinear differential equations. In their mean field approach, nonlinear evolution of a quantum state $\KET{\varphi}$ is generated through coupling to many identical, weakly interacting copies of $\KET{\varphi}$, as in a Bose-Einstein condensate. In quantum many-body models for $n$ indistinguishable atoms satisfying Bose statistics and prepared in a product state, it has been rigorously established that the nonlinear Gross-Pitaevskii equation for the 1-particle density matrix becomes exact in the large $n$  or thermodynamic limit, i.e., the 1-particle nonlinear Gross-Pitaevskii equation is {\it dual} to the $n$-particle linear Schr\"odinger equation when $n \rightarrow \infty$ \cite{SpohnRMP1980,BardosMAA2000,200605486,FrohlichCMP07,ErdosPRL2007,RodnianskiCMP09,ErdosJSP09,KnowlesCMP10,ChenLeeJSP11,14116284,PicklRMP15,Benedikter2016,FrohlichAM19,BrenneckeAPED19,191014521}. As with bosons, and some spin models \cite{SpohnRMP1980,EllisLargeDeviations1985}, the mean field approach of Ref.~\cite{201106571} is also expected to become exact in the large $n$ limit, but the precise form of this convergence has not been determined.

Here we extend the linear/nonlinear duality to $n$ qubits subjected to arbitrary 1-qubit and SWAP-symmetric 2-qubit unitaries, a generalized central ``spin" model (CSM) \cite{GaudinJPF1976,YuzbashyanAltshulerJPA2005,BortzPRB2010,HeChesiPRB2019,220101025,9511011,BortzJSM2007,220909225,KesslerPRA2012,MukhopadhyayBhattacharyaPRA2017}. 
The objectives are as follows: (i) Use mean field theory to construct a rigorous duality between nonlinear qubits and a many-body CSM evolving under standard linear quantum mechanics. (ii) Provide an upper bound for the {\it model} error associated with the use of mean field theory, and investigate its breakdown at large times. (iii) Highlight the origin of qubit 
torsion, 
which leads to {\it expansive} dynamics, where the trace distance between a pair of close qubit states increases with time \cite{MielnikJMP80,PhysRevLett.81.3992,BechmannPLA98,0502072,150706334,211105977}. Section \ref{central spin model section} defines the CSM. Section \ref{large n limit section} employs the proof techniques of \cite{ErdosJSP09} and \cite{NachtergaeleJSP06} to establish the duality. Section \ref{discussion section} explains the origin of qubit torsion within this framework, and contains the conclusions. Simulated examples and additional information are provided in an appendix.
 
\section{Central spin model}
\label{central spin model section}

\subsection{Model definition}

Let $\{ 1, 2, \cdots , n \}$ denote the vertices of a star graph of $n$ qubits. Qubit 1 is the {\bf central qubit}, and the remaining ancilla qubits $\{ 2, \cdots , n\}$ are used to simulate a certain type of environment for the central qubit. However this simulated environment is far from that of a random, noisy bath. Instead, the ancilla qubits are initialized in the same pure state $\KET{\varphi}$ and they couple symmetrically to the central qubit.  We consider a generalized homogeneous CSM with Hamiltonian
\begin{eqnarray}
 H = \sum_{i=1}^n H^0_i + \frac{1}{n \! - \! 1} \bigg( \sum_{j>1}^n V_{1j} + \lambda \sum_{i>1}^{n-1} \sum_{j > i}^n V_{ij} \bigg), \ \ 
 [ V_{ij} , \chi_{ij} ] = 0,
 \ \  -1 \le \lambda \le 1.
 \label{central spin model}
 \end{eqnarray}
The  Hamiltonian $H^0_i$ acts as  $H^0 \in \mathfrak{su}(2)$ on qubit $i$ and as the identity otherwise. Each qubit $i \in \{ 1, 2, \cdots, n\}$ sees the same single-qubit Hamiltonian $H^0$. 
This can be further expanded in a basis of Pauli matrices as $H^0_{i} = \! \! \sum_{\mu = 1}^3 B_{\mu} \, \sigma_i^\mu $, where the ``field" $ {\vec B} = (B_1, B_2, B_3) \in {\mathbb R}^3$ has no dependence on the qubit index $i$. 
Interaction $V_{ij}$  acts as $V \in \mathfrak{su}(4)$ on the edge $(i,j)$ and as the identity otherwise. In addition, we require $V_{ij}$ to be SWAP-symmetric, where SWAP is a two-qubit operator that acts on a product state as $ \chi_{ij}   \KET{\alpha}_i \! \otimes \! \KET{\beta}_j = \KET{\beta}_i \! \otimes \!  \KET{\alpha}_j $. 
Note that the interaction in (\ref{central spin model}) has infinite range, favoring a mean field description.  A factor $O(1/n)$ is needed to control the large $n$ limit and is typical in large $n$ problems.

The parameter $ \lambda $ controls the ancilla-ancilla coupling and therefore affects the permutation symmetry of the Hamiltonian. We are mainly interested in $\lambda = 0$ but also consider cases with  $ |\lambda|  \le 1$. A CSM with $\lambda \neq 0$ might apply to two species of atomic qubits with inhomogeneous interactions. The case $\lambda =1$ applies when all qubits are symmetrically coupled and the interaction graph is complete. Call this the {\bf complete graph} (CG) model:
\begin{eqnarray}
H_{\rm CG} = \sum_{i=1}^n H^0_i + \frac{1}{n \! - \! 1} \sum_{i=1}^{n-1} \sum_{j > i}^n V_{ij} .
\label{cg model}
\end{eqnarray}
The CG model (\ref{cg model}) is a qubit analog of a weakly interacting monatomic Bose gas. Although we treat it  as a special case of the CSM, they are distinct models with different  symmetries. 

\clearpage

A general SWAP-symmetric interaction can be obtained from the Cartan decomposition  of  $\mathfrak{su}(4)$ \cite{KhanejaCP01}, with which any $U \in {\rm SU(4)}$ can be written as an element of ${\rm SU}(2)_i \otimes {\rm SU(2)_j }$, followed by a symmetric entangling gate 
$ e^{- \I \sum_\mu J_\mu \sigma_i^\mu \otimes \sigma_j^\mu}$, 
then a second ${\rm SU}(2)_i \, \otimes  \, {\rm SU(2)_j}$. SWAP symmetry requires that the SU(2) unitaries in $V_{ij}$ are the same on every qubit. They can therefore be generated by a single-qubit Hamiltonian $H^0$ and are not explicitly included in the interaction, which then takes the form 
\begin{eqnarray}
V_{ij} = \sum_{\mu = 1}^3 J_\mu \, \sigma_i^\mu \otimes \sigma_j^\mu \! , \ \ \ {\vec J} = (J_1, J_2, J_3) \in {\mathbb R}^3,
\label{interaction pauli expansion}
\end{eqnarray}
where the couplings $J_{\mu}$ have no dependence on the edge label $(i,j)$. The qubits interact via a {\it vector} coupling and have three coupling constants $J_1, J_2, J_3$, instead of one as in the monatomic Bose gas case. 

The operators $H_i^0$ and $V_{ij}$ are time-dependent and subject to the conditions that the quantities 
\begin{eqnarray}
\nu_0 :=  \sup_t  \| H_i^0(t) \|_\infty
\ \ \ {\rm and} \ \ \ 
J_0 := \sup_{\mu, t}  |J_\mu(t) | ,
\label{interaction bound}
\end{eqnarray}
exist and are finite. Here $ \| \cdot \|_\infty $ is the operator norm (relevant norm properties are collected in the appendix). The quantity $J_0 $ bounds the coupling, and hence the buildup of multiqubit correlation and corresponding breakdown of mean field theory. 

The time-evolution operator for the CSM is
 \begin{eqnarray}
 U_{t} = T e^{- \I \int_0^t \! H(\tau) \, d\tau} , \ \ 
\frac{ d U_{t}  }{dt}  = - \I  H(t) \, U_{t}, \ \ 
U_{0} = I,
\label{central spin model evolution}
 \end{eqnarray}
where  $T$ is the time-ordering operator, $I$ is the identity, $\I = \sqrt{-1}$, and factors of $\hbar$ are suppressed throughout this paper. 
We will also need the time-evolution operator for any single uncoupled qubit, which is
\begin{eqnarray}
u_{t} = T e^{-\I  \int_0^t \! H^0(\tau) \, d\tau} \! , \ \
\frac{d u_t}{dt} = -\I H^0(t) \, u_{t}, \ \  u_{0} = I.\label{free qubit evolution}
\end{eqnarray}

\clearpage

The CSM with $\lambda \! = \! 0$  has a long history and many variants have been investigated \cite{GaudinJPF1976,YuzbashyanAltshulerJPA2005,BortzPRB2010,HeChesiPRB2019,220101025,9511011,BortzJSM2007,220909225,KesslerPRA2012,MukhopadhyayBhattacharyaPRA2017}. Models with XXX symmetry [by which we mean $ {\vec J}  \! = \! (J_1,J_1,J_1)$]
and some with XXZ symmetry 
$[{\vec J} \! = \!  (J_1,J_1,J_3)]$ are integrable and exactly solvable by Bethe ansatz \cite{GaudinJPF1976,YuzbashyanAltshulerJPA2005,BortzPRB2010,HeChesiPRB2019,220101025}.  The $\lambda \! = \! 0$ CSM with Heisenberg interaction, XXX, has been studied extensively \cite{GaudinJPF1976,YuzbashyanAltshulerJPA2005,BortzPRB2010,HeChesiPRB2019,220101025,9511011,BortzJSM2007,220909225}. Time-dependent mean field solutions in the XXX case  have been obtained in terms of hyperelliptic functions \cite{YuzbashyanAltshulerJPA2005}.
Phase transitions have also been studied  \cite{KesslerPRA2012,MukhopadhyayBhattacharyaPRA2017}. In this paper we study solutions of the CSM with XYZ interaction 
[arbitrary 
bounded ${\vec J} \! = \!  (J_1,J_2,J_3)], $ 
general $ \lambda $,
 and high degrees of permutation symmetry. 
Specifically, we consider two levels of permutation symmetry:
 \begin{enumerate}
 
\item[$S_{n-1}$:] 
This is the symmetry of the $\lambda \neq 1$ model, which includes the set of all permutations among ancilla  $\{ 2, \cdots , n \}$. The symmetry group of the model then contains a subgroup of the symmetric group $S_n$ (permutations on $n$ qubits) that we simply call $S_{n-1}$. 

\item[$S_{n}$:]
 The higher symmetry case has full permutation symmetry, including the central qubit. This is the symmetry of the  $\lambda = 1$ model. Now the symmetry group contains $S_n$. 
\end{enumerate}
We note that the initial condition
$ \rho(0)$ will respect both symmetries.

\subsection{Linear picture: BBGKY hierarchy} 
\label{linear picture section} 

At time $t=0$ the central qubit and ancilla are prepared in a product state 
\begin{eqnarray}
\rho(0) = \KETBRA{\varphi}^{\otimes n} \! \! , \ \KET{\varphi} = \varphi_0 \KET{0} +  \varphi_1 \KET{1}, \  \varphi_{0,1} \in {\mathbb C}, \ 
|\varphi_0|^2 +  |\varphi_1|^2  = 1. 
\end{eqnarray}
This initial condition has complete permutation symmetry, $S_n$.  At later times $ t > 0$ the state is $\rho(t) = U_{t} \, (\KETBRA{\varphi}^{\otimes n}) \, U_{t}^\dagger$ and the evolution equation is
 \begin{eqnarray}
 \frac{d \rho}{dt} = -\I \big[ \sum_{i=1}^n H^0_i , \rho \big]
 - \I  \big[  \sum_{j>1}^n \frac{V_{1j}}{n-1} + \lambda \sum_{i>1}^{n-1} \sum_{j > i}^n \frac{V_{ij}}{n-1}   , \, \rho \, \big], \ \ \ -1 \le \lambda \le 1.
 \end{eqnarray}
Let  $ {\rm tr}_{i}(\cdot) = \sum_{x=0,1} \BRA{x} \cdot \KET{x}_{i} $ denote the partial trace over the Hilbert space of qubit $ i $.
 The density matrix for the central qubit is 
 $ \rho_1(t) = {\rm tr}_{>1}[\rho(t)] $,
 where
$ {\rm tr}_{>i}(\cdot) := {\rm tr}_{i+1} {\rm tr}_{i+2}  \cdots {\rm tr}_{n} (\cdot)$.
Similarly, $ \rho_2(t) = {\rm tr}_{1}[\rho_{12}(t)] $,
where 
$ \rho_{12} =  {\rm tr}_{>2}[\rho(t)]. $ 
Then we have
\begin{eqnarray}
 \frac{d \rho_1}{dt} &=& -\I [ H^0 \! , \, \rho_1 ]
 - \I  \, {\rm tr}_{>1}   \big[  \sum_{j>1}^n \frac{V_{1j}}{n \! - \! 1} + \lambda \sum_{i>1}^{n \! - \! 1} \sum_{j > i}^n \frac{V_{ij}}{n \! - \! 1}   , \, \rho \, \big] \\
 &=& -\I [ H^0 \! , \, \rho_1 ]
 - \I  \, {\rm tr}_{>1}   \big[  \sum_{j>1}^n \frac{V_{1j}}{n \! - \! 1}  , \rho \, \big] , \\
 \frac{d \rho_2}{dt} &=& -\I [ H^0 \!  , \, \rho_2 ]
 - \I  \, 
 {\rm tr}_{1} {\rm tr}_{3} \cdots {\rm tr}_{n}   \big[  \sum_{j>1}^n \frac{V_{1j}}{n \! - \! 1} + \lambda \sum_{i>1}^{n \! - \! 1} \sum_{j > i}^n \frac{V_{ij}}{n \! - \! 1}   , \, \rho \, \big] \\
 &=&
 -\I [ H^0 \!  , \, \rho_2 ]  - \I  \, 
 {\rm tr}_{1} {\rm tr}_{3} \cdots {\rm tr}_{n}   \big[ \frac{V_{12}}{n \! - \! 1} + \lambda \sum_{j > 2}^n \frac{V_{2j}}{n \! - \! 1}   , \, \rho \, \big] ,
 \end{eqnarray}
 using (\ref{first partial trace identity}) and 
 (\ref{second partial trace identity}).
 Next we assume $S_{n-1}$ ancilla permutation symmetry to obtain
 \begin{eqnarray}
 \frac{d \rho_1}{dt} 
 &=& -\I \, [ H^0 \! , \, \rho_1 ]
 - \I \, {\rm tr}_{2} ( [V_{12} ,  \rho_{12} ]). \nonumber \\
 &=& -\I \sum_{\mu=1}^{3}  B_\mu \, [ \sigma_1^\mu   , \, \rho_1 ] 
 - \I  \sum_{\mu=1}^{3} J_\mu \,  
   [ \sigma_1^\mu , \, {\rm tr}_{2}(\rho_{12}  \sigma_2^\mu )  ]. 
 \label{rho1 evolution equation} \\
  \frac{d \rho_2}{dt}  &=&
 -\I [ H^0 \!  , \, \rho_2 ]
 - \I  \, \frac{ {\rm tr}_{1} [ V_{12}, \rho_{12} ]  + \lambda (n-2) \, {\rm tr}_{3}  \,  [ V_{23}   , \, \rho_{23} \, ] }{n-1} \nonumber \\
 &=&
 -\I \sum_{\mu=1}^{3} B_\mu \, [ \sigma_2^\mu   , \, \rho_2 ] 
 - \I \sum_{\mu=1}^{3} \frac{J_\mu}{n \! - \! 1}    
\big [ \sigma_2^\mu  , \,   {\rm tr}_{1}(\rho_{12} \sigma_1^\mu) + \lambda (n-2) \,  {\rm tr}_{3}(\rho_{23} \sigma_3^\mu) \big] ,
 \label{rho2 evolution equation}
 \end{eqnarray}
 where ${\vec B}$ and ${\vec J}$ are possibly time-dependent.  From these we obtain
\begin{eqnarray}
\rho_1(t) &=& u_t \bigg( \! \! \KETBRA{\varphi} 
- \I  \sum_\mu  \int_0^t \! d\tau \, J_\mu  \, u^\dagger_{\tau}  
\, [ \sigma_1^\mu , \, {\rm tr}_{2}(\rho_{12}  \sigma_2^\mu )  ] \, u_{\tau} \!
\bigg) u_t^\dagger 
 , \nonumber 
 \\
\rho_2(t) &=& u_t \bigg( \! \! \KETBRA{\varphi} 
- \I  \sum_\mu  \int_0^t \! d\tau \, \frac{J_\mu}{n \! - \! 1}  \, u^\dagger_{\tau}  
\, [ \sigma_2^\mu , \,  {\rm tr}_{1}(\rho_{12} \sigma_1^\mu) + \lambda (n-2) \,  {\rm tr}_{3}(\rho_{23} \sigma_3^\mu)  ] \, u_{\tau} \!  \bigg) u_t^\dagger,  \ \ 
\label{linear integral equations}
\end{eqnarray}
where $\rho_{23} = {\rm tr}_{1} (\rho_{123}) = {\rm tr}_1( {\rm tr}_{>3} \rho) $. 
Here $ u_{t}$ is the time-evolution operator
(\ref{free qubit evolution})
 for a single uncoupled qubit. 
 The equations for $\rho_{1,2}$ are
 quantum Bogoliubov-Born-Green-Kirkwood-Yvon (BBGKY) hierarchy equations \cite{FesciyanCMP73} for the generalized CSM. 

\subsection{Nonlinear picture: Mean field theory} 

Theorem \ref{erdos-schlein theorem} in 
Sec. \ref{large n limit section} relates the solutions of (\ref{rho1 evolution equation}-\ref{rho2 evolution equation}) to that of a mean field theory model. To construct that model, assume that the {\bf order parameter}
\begin{eqnarray}
{\vec m}_i :=  \langle {\vec \sigma}_i  \rangle = {\rm tr}(\omega {\vec \sigma}_i ), \ \ i \in \{1,2, \cdots, n\}
\end{eqnarray}
is nonvanishing, where the expectation is with respect to some (possibly time-dependent) state $\omega$. To find equilibrium properties, $\omega$ is assumed to be a thermal state $e^{- \beta H} /( {\rm tr} \, e^{- \beta H} ) $ at temperature $1/\beta$. Here we assume that  $\omega$ is arbitrary (to be specified) and time-dependent. Expanding the Hamiltonian (\ref{central spin model}) in powers of fluctuations 
$ \delta \sigma_i^\mu  = \sigma_i^\mu \! - \! m_i^\mu$ to first order gives
\begin{eqnarray}
H &=& \sum_{i=1}^n H^0_i + 
  \sum_\mu \frac{J_\mu}{n \! - \! 1}  
 \sum_{j>1}^n 
 \big(    m_1^\mu  \, \sigma_j^\mu  
  \! + \!  \sigma_1^\mu  \, m_j^\mu \big)  
  \nonumber \\
 &+& \lambda   \sum_\mu \frac{J_\mu}{n \! - \! 1} 
 \sum_{i>1}^{n-1} \sum_{j>i}^n 
 \big( m_i^\mu  \, \sigma_j^\mu  
 \! + \!  \sigma_i^\mu  \,  m_j^\mu   
 \big) + \Delta E,
 \label{general inhomogeneous mft}
\end{eqnarray}
where
\begin{eqnarray}
 \Delta E = - \sum_\mu \sum_{i>1}^n
 \frac{J_\mu \, m_1^\mu  m_i^\mu  }{n \! - \! 1}  
- \lambda \sum_\mu  \sum_{i>1}^{n-1} \sum_{j>i}^n   \frac{J_\mu  \, m_i^\mu m_j^\mu  }{n \! - \! 1}  .
\label{general inhomogeneous background}
\end{eqnarray}
The ``background" energy  $\Delta E$ has no affect on the dynamics but contributes to thermodynamic properties such as the free energy.

In the following section we construct a mean field theory for CSM solutions with $S_{n-1}$ symmetry. The result is  a pair of coupled equations of motion for the mean field state $X$ of the central qubit, and the mean field state $Y$ of an ancilla (qubit 2). Because the equations of motion are coupled, they must be  solved together. Hence, the dual mean field model is a two-qubit model in a separable state $X \otimes Y$. This is the primary mean field theory for the CSM.
An exception occurs if $\lambda = 1$: In this case, assuming $X(0)=Y(0) = \KETBRA{\varphi}$, the coupled equations of motion yield $X(t)=Y(t)$ for all time, leading to a solution with $S_{n}$ symmetry. The mean field theory for this case is also discussed below. The CSM with $\lambda = 1$ preserves the $S_{n}$ symmetry of the initial condition, leading to a  single-qubit dual model with self interaction.

\clearpage

\subsubsection{Symmetry $S_{n-1}$}
\label{sn-1 symmetry section}

If the CSM exhibits $S_{n-1}$ symmetry, the order 
parameter satisfies $ {\vec m}_2 = {\vec m}_3 = \cdots = {\vec m}_n$. Then from (\ref{general inhomogeneous mft}) we obtain
\begin{eqnarray}
 H &=& \sum_{i=1}^n H^0_i +  \sum_\mu J_\mu \,  m_2^\mu \, \sigma_1^\mu + \sum_\mu   \frac{ J_\mu  m_1^\mu }{n\! - \! 1} 
 \sum_{i>1}^n \sigma_i^\mu 
 + \lambda  \sum_\mu \frac{J_\mu m_2^\mu }{n \! - \! 1}  \sum_{i>1}^{n-1} \sum_{j>i}^n ( \sigma_i^\mu + \sigma_j^\mu)
 + \Delta E \nonumber \\
  &=& \sum_{i=1}^n H^0_i +  \sum_\mu J_\mu \,  m_2^\mu \, \sigma_1^\mu + \sum_\mu  \frac{ J_\mu  m^\mu_1 + \lambda (n-2) J_\mu  m^\mu_2 }{n-1} 
 \sum_{i>1}^n \sigma_i^\mu  + \Delta E,
\end{eqnarray}
where
\begin{eqnarray}
\Delta E = - \sum_\mu J_\mu  \, m_1^\mu  m_2^\mu 
- \frac{\lambda}{2} \sum_\mu (n-2) J_\mu m_2^\mu m_2^\mu .
\label{background energy}
 \end{eqnarray}
In the mean field approximation (neglecting quadratic fluctuations) the qubits are decoupled and the mean field Hamiltonians for qubits 1 and 2 are 
\begin{eqnarray}
H^{\rm eff}_{1} &=& H^0 +\sum_\mu J_\mu \,
{\rm tr}(Y \sigma^\mu) \, \sigma_1^\mu \\
H^{\rm eff}_{2} &=& H^0 +\sum_\mu J_\mu \frac{ {\rm tr}(X \sigma^\mu) \! + \!  \lambda (n \! - \! 2)  {\rm tr}(Y \sigma^\mu) }{n \! - \! 1}  \,\sigma_2^\mu ,
\end{eqnarray}
where $X$ and $Y$ are the mean field density matrices for qubits 1 and 2, respectively. Here we have set $\omega = X \otimes Y$, the current mean field state of qubits 1 and 2. The evolution equations for $X$ and $Y$ are
\begin{eqnarray}
\frac{dX}{dt} &=& - \I [H^0, X] - \I  \sum_{\mu=1}^{3}
J_\mu \,
{\rm tr}(Y  \sigma^\mu) \, [ \sigma^\mu, X],  
\label{x evolution equation} \\
 \frac{dY}{dt}  &=&  - \I [H^0, Y] - \I 
\ \sum_{\mu=1}^{3} J_\mu \,
\frac{ {\rm tr}(X \sigma^\mu) \! + \!  \lambda (n \! - \! 2)  {\rm tr}(Y \sigma^\mu) }{n \! - \! 1} \,
[ \sigma^\mu, Y] , 
\label{y evolution equation}  \\
& \approx & - \I [H^0, Y] - \I \, \lambda \!
\ \sum_\mu J_\mu \,
 {\rm tr}(Y \sigma^\mu) \, [ \sigma^\mu, Y]  ,
\label{y evolution equation large n} 
 \end{eqnarray}
 where  (\ref{y evolution equation large n}) applies in the large $n$ limit. 
The initial conditions are 
 \begin{eqnarray}
 X(0)  = Y(0) = \KETBRA{\varphi}.
 \end{eqnarray}
Next, using (\ref{free qubit evolution}), we obtain
\begin{eqnarray}
X(t) &=& u_t \bigg( \! \! \KETBRA{\varphi} - \I 
\sum_\mu  \!  \int_0^t \! d\tau \, J_\mu \,  {\rm tr}(Y  \sigma^\mu)  \, u_{\tau}^\dagger   \big( [ \sigma^\mu, X]  \big) u_{\tau} \! \bigg) u_t^\dagger 
 , \nonumber \\
Y(t) &=& u_t \bigg( \! \! \KETBRA{\varphi} - \I \sum_\mu \! \int_0^t \! d\tau \, J_\mu \,  
\frac{ {\rm tr}(X \sigma^\mu) \! + \!  \lambda (n \! - \! 2) \,  {\rm tr}(Y \sigma^\mu) }{n \! - \! 1} \,
\, u_{\tau}^\dagger  \big( [ \sigma^\mu, Y]  \big) u_{\tau} \! \bigg) u_t^\dagger \ \ \ \ \ 
\label{nonlinear integral equations}
\end{eqnarray}
The nonlinear evolution equations (\ref{x evolution equation}) and (\ref{y evolution equation}) are dual to the linear BBGKY equations (\ref{rho1 evolution equation}) and (\ref{rho2 evolution equation}) in the large $n$ limit in the sense that $X = \rho_1$ and $Y = \rho_2$  in this limit. This is because Theorem \ref{erdos-schlein theorem} implies 
$\lim_{n \rightarrow \infty} \| X - \rho_1 \| \rightarrow 0$ and $\lim_{n \rightarrow \infty} \| Y - \rho_2  \| \rightarrow 0$. 

\subsubsection{Symmetry $S_{n}$}
\label{sn symmetry section}

If the CSM exhibits $S_{n}$ symmetry, the order 
parameter satisfies $ {\vec m}_1 = {\vec m}_2 = \cdots = {\vec m}_n$. For $ {\vec m}_1 $ and  $ {\vec m}_2 $ to be equal, we must have 
$X=Y$,\footnote{This is because, for a qubit, the order parameter $ {\vec m} = {\rm tr}( \rho {\vec \sigma}) $ uniquely specifies the state $ \rho = (I + {\vec m} \cdot {\vec \sigma})/2.$} 
indicating symmetry between the central and ancilla qubits. Here we use the mean field equations (\ref{x evolution equation}) and (\ref{y evolution equation}) to investigate $S_{n}$ symmetry as a special case of $S_{n-1}$ symmetry. First transform to
\begin{eqnarray}
\rho_{\rm ave}:= \frac{X+Y}{2} \ \ {\rm and} \ \  
\rho_{\! \Delta} := \frac{X-Y}{2} .
 \end{eqnarray}
 While $\rho_{\rm ave}$ is a state (positive semidefinite matrix with unit trace), $\rho_{\! \Delta} $ is not. 
For large $n$,
\begin{eqnarray}
\frac{d \rho_{\rm ave}}{dt} &=&  - \I [H^0, \rho_{\rm ave}] - \I \, 
\sum_\mu J_\mu \,
 {\rm tr}(\rho_{\rm ave} \sigma^\mu \! - \!  \rho_{\! \Delta}  \sigma^\mu)  \, \big[ \sigma^\mu,  \rho_{\rm ave}+ (\lambda \! - \! 1) \frac{\rho_{\rm ave} \! - \!  \rho_{\! \Delta} }{2}   \big] , \\
 \frac{d\rho_{\! \Delta}  }{dt} &=&   - \I [H^0, \rho_{\! \Delta}  ] - \I \, 
\sum_\mu J_\mu \,  {\rm tr}(\rho_{\rm ave} \sigma^\mu \! - \!  \rho_{\! \Delta}  \sigma^\mu)   \, \big[ \sigma^\mu,  \rho_{\! \Delta}  +  (1 \! - \! \lambda)  \frac{\rho_{\rm ave} \! - \!  \rho_{\! \Delta} }{2}  \big] ,
 \end{eqnarray}
 with initial conditions
 $ \rho_{\rm ave}(0) = \KETBRA{\varphi} $
 and
$ \rho_{\! \Delta} (0) = 0.$
At time zero, $\rho_{\! \Delta} =0$, so the system initially possesses $S_{n}$ symmetry. If $\lambda \neq 1$, the initial rate of change 
$ (d\rho_{\! \Delta} /dt)_0 =  - \I  (\frac{1 - \lambda}{2})
\sum_\mu J_\mu \, {\rm tr}( \rho_{\rm ave}\sigma^\mu)  \, [ \sigma^\mu  , \rho_{\rm ave}   ] $ is nonzero, breaking the symmetry between $X$ and $Y$. However $\rho_{\! \Delta} $ remains zero if $\lambda = 1$, preserving the $S_{n}$ symmetry and leading to a single-qubit mean field theory for $X$ with self-interaction:
\begin{eqnarray}
\frac{dX}{dt}  &=&  - \I \, [H^0, X ] - \I \, 
\sum_{\mu=1}^{3} J_\mu \, {\rm tr}(X \sigma^\mu)  \, [ \sigma^\mu,  X] .
\label{self interaction model}
\end{eqnarray}

\clearpage

\section{Large $n$ limit}
\label{large n limit section}

In this section we establish the duality between the linear BBGKY equations and the nonlinear mean field theory in the large $n$ limit of the generalized CSM, following the proof techniques of \cite{ErdosJSP09} and \cite{NachtergaeleJSP06}.
Our work also builds on recent papers by Fernengel and Drossel \cite{190709349} and K{\l}obus {\it et al.}~\cite{211113477} who studied 
nonlinear mean field dynamics of related spin models. Some features of our analysis are:
 (1) In contrast to particle models, we do not assume indistinguishable particles with Bose or Fermi statistics. (2) The $\lambda = 0$ model has reduced permutation symmetry and no interaction between ancilla. Full permutational symmetry is broken, but the ancilla qubits $\{ 2, \cdots , n \}$ remain identical. (3) Qubits interact via an arbitrary $V \in \mathfrak{su}(4)$. (4) The interaction is long ranged and does not decay with distance. (5) All terms in the Hamiltonian are assumed to be time dependent.

\begin{thm}[Extended Erd\H{o}s-Schlein \cite{ErdosJSP09}]
Let $X(t)$ and $Y(t)$ be solutions of the coupled nonlinear evolution equations (\ref{x evolution equation}) and (\ref{y evolution equation}) [or  (\ref{y evolution equation large n})]  for the $n$-qubit generalized CSM (\ref{central spin model}), with initial conditions $X(0) = Y(0) = \KETBRA{\varphi}$,  where $\KET{\varphi} = \varphi_0 \KET{0} +  \varphi_1 \KET{1}$,  $\varphi_{0,1} \in {\mathbb C}$, $|\varphi_0|^2 +  |\varphi_1|^2  \! = \! 1$. 
Also let  $ \rho_1 = {\rm tr}_{>1}(\rho) $ and 
$ \rho_{2} = {\rm tr}_{1}(\rho_{12}) $  be the exact reduced density matrices on qubits 1 and 2, respectively (partial trace notation is defined in Sec.~\ref{linear picture section}).
Then the distance in trace norm between the mean field and exact state satisfies
\begin{eqnarray}
\| X(t) - \rho_1(t) \|_1 \le 
 4 \,  \frac{e^{ 12 (1+  |\lambda| ) J_0 t }-1 }{n (1+  |\lambda| ) }, 
 \ \ t \ge 0, 
 \label{erdos-schlein thm x equation}
 \end{eqnarray}
 and
 \begin{eqnarray}
 \| Y(t) - \rho_2(t) \|_1 \le 
 4 \,  \frac{e^{ 12 ( 1+  |\lambda| ) J_0 t }-1 }{n (1+  |\lambda| ) }, 
 \ \ t \ge 0, 
\label{erdos-schlein thm y equation}
\end{eqnarray}
where $ J_0$ is an interaction strength bound defined in (\ref{interaction bound}). The same upper bound applies to both $X$ and $Y$. The inequalities imply that, 
for any fixed $t \ge 0$,
\begin{eqnarray}
 \lim_{n \rightarrow \infty} \| X(t) - \rho_1 (t) \|_1 = 0,  \\
  \lim_{n \rightarrow \infty} \| Y(t) - \rho_2 (t) \|_1 = 0, 
\label{erdos-schlein thm limit equation}
\end{eqnarray}
establishing the duality.

\label{erdos-schlein theorem}
\end{thm}

\clearpage

\noindent The proof of Theorem \ref{erdos-schlein theorem} uses the following lemmas:

\begin{lem}[Lieb-Robinson Bound \cite{LiebCMP72,ErdosJSP09}] 
For any $k \in \{ 1, \cdots, n-1 \}$,  let 
$A_{1, \dots , k} \in \CMATRIX{2^n}$ and 
$B_{k+1} \in \CMATRIX{2^n}$ 
be Hermitian bounded linear operators 
(observables) with support exclusively in subsets $\{ 1,2, \cdots , k \}$ and $\{ k + 1 \} $, respectively,  of the $n$-qubit generalized CSM (\ref{central spin model}). Here $A_{1, \dots , k}$ acts nontrivially on the first $k$ qubits $\{ 1,2, \cdots , k \} $ (including the central qubit) and as the identity elsewhere. Similarly, $B_{k+1}$ acts nontrivially on qubit $k+1$ only. Let
\begin{eqnarray}
\Gamma_{\! kt} := \! \!  \sup_{A \neq 0 , B\neq 0} \frac{ \| [ U_{t}^\dagger  A_{1, \dots , k} U_{t}  , B_{k+1} ] \|_\infty  }{ \| A_{1, \dots , k} \|_\infty \,  \| 
B_{k+1} \|_\infty } ,
\label{lieb-robinson bound}
\end{eqnarray}
where the supremum is over the set of all bounded linear operators $A_{1, \dots , k}$ with support on qubits $\{ 1, \cdots, k\}$ such that $ \| A_{1, \dots , k} \|_\infty \neq 0$, and  over all bounded linear operators $B_{k+1}$ with support on qubit $k+1$ such that $ \| B_{k+1} \|_\infty \neq 0$. 
Then
\begin{eqnarray}
\Gamma_{\! kt} \le 2
\label{lemma 1 upper bound}
\end{eqnarray}
holds for any $k=  1, 2, \cdots, n-1$. Furthermore,
for $k=1,2$,
\begin{eqnarray}
\Gamma_{\! kt}    \le  
2 \, \frac{e^{6 (1 \! + \! |\lambda|) J_0 \,  t } - 1  }{ n-1} ,
\label{lemma 1 equation}
\end{eqnarray}
where $J_0$ is defined in 
(\ref{interaction bound}).
\label{lemma 1}
\end{lem}

The quantity $\Gamma_{\! kt}$ is a measure of the largest possible correlation between a cluster containing the first $k$ qubits (including the central), and qubit $k+1$, due to their interaction. 
Only cases $k=1,2$ are required below. The bound (\ref{lemma 1 upper bound}) shows  that correlation measured this way does not blow up at long times, in contrast with (\ref{lemma 1 equation}). Therefore the interesting regime  occurs when the bound in (\ref{lemma 1 equation}) is small, namely 
$ n \gg e^{6 (1+ |\lambda|) J_0 t}$. 

\vspace{0.2 in}

\begin{proof}
The bound (\ref{lemma 1 upper bound}) follows from unitary invariance and 
submultiplicativity of the Schatten $p$-norm (see appendix).  To obtain (\ref{lemma 1 equation}), transform to a representation where time-evolution is generated exclusively by the cross-interactions
 \begin{eqnarray}
W^{(k)}  &:=& \frac{1}{n-1} \bigg( \sum_{j=k+1}^n V_{1j}
+ \lambda \sum_{i=2}^k \sum_{j=k+1}^n V_{ij}
\bigg) 
\label{w definition} 
\end{eqnarray}
between the $k$-qubit cluster on which $A_{1, \dots , k}$ acts, and its environment.
In particular,
\begin{eqnarray}
W^{(k=1)}  =  \frac{ V_{12} }{ n-1 }
+  \frac{V_{13} + \cdots + V_{1n}}{n-1}, 
\end{eqnarray}
independent of $\lambda$, and
\begin{eqnarray}
W^{(k=2)}  =  \frac{V_{13} +  \lambda V_{23} }{n-1}
+  \frac{V_{14} + \cdots + V_{1n}
+ \lambda (V_{24} + \cdots + V_{2n} ) }{n-1}  . \end{eqnarray}
In these expressions, terms that don't commute with $B_{k+1}$ have been isolated.
The first step of the proof is to note that
\begin{eqnarray}
\frac{d}{dt} (U_{t}^\dagger  S_{kt} A_{1, \dots , k} S_{kt}^\dagger U_{t} ) 
 = \I \, [ U_{t}^\dagger W^{(k)} U_{t} , U_{t}^\dagger  S_{kt} A_{1, \dots , k} S_{kt}^\dagger U_{t}  ]
 = \I \, [  \WFRAME{k}  \! , U_{t}^\dagger  S_{kt} A_{1, \dots , k} S_{kt}^\dagger U_{t}  ], 
 \end{eqnarray}
 where, for any $k \in \{ 1,2, \cdots, n-1 \}$, 
 \begin{eqnarray}
 H^{(k)}  &=&  H -  W^{(k)}, 
 \ S_{kt} = T e^{- \I \int_0^t \! H^{(k)}(\tau) \, d\tau} , \  \frac{ d S_{kt}  }{dt}  \! = \!  - \I H^{(k)}(t) \, S_{kt},  \ \  S_{k0}= I,   \\
 \WFRAME{k} &=& U_{t}^\dagger W^{(k)} U_{t}, \ \ 
\SFRAME{kt} = T e^{ \I \int_0^t   \WFRAME{k}(\tau) \, d\tau} , \ \ 
\frac{d \,  \SFRAME{kt} }{dt} =  \I \,  \WFRAME{k}(t) \, \SFRAME{kt}, \ \ \SFRAME{k0} = I .
 \end{eqnarray}
 The time-evolution operators
$ S_{kt}$ and $\SFRAME{kt} $ are generated by 
$- \I  {H}^{(k)}$ and $ \I \WFRAME{k}  \! , $ respectively.
Hamiltonian ${H}^{(k)}$ has the cross-interactions $W^{(k)}$ between the $k$-qubit cluster and its surroundings removed. 
Next let $f_{kt} := [ U_{t}^\dagger  S_{kt} A_{1, \dots , k} S_{kt}^\dagger U_{t} , B_{k+1} ]$. 
 Then
\begin{eqnarray}
\frac{df_{kt}}{dt} &=& \I \,  \big[  
[  \WFRAME{k} , U_{t}^\dagger  S_{kt} A_{1, \dots , k} S_{kt}^\dagger U_{t}  ], B_{k+1} \big] = 
\I \, [  \WFRAME{k} , \, f_{kt} ] + c_{kt}, 
\end{eqnarray}
 where
 $ c_{kt} = \I \, \big[ [    \WFRAME{k} , \, B_{k+1} ] ,   U_{t}^\dagger S_{kt} A_{1, \dots , k} S_{kt}^\dagger  U_{t} \big].$  
 We then have $\frac{d}{dt} (\SFRAME{kt}^\dagger f_{kt} \SFRAME{kt}) =  \SFRAME{kt}^\dagger c_{kt} \SFRAME{kt}$
 and
$ \SFRAME{kt}^\dagger f_{kt} \SFRAME{kt}   = 
\int_{0}^t   \SFRAME{k \tau}^\dagger c_{k \tau} \SFRAME{k\tau} \, d\tau$,
because 
$f_{k0} =  [ A_{1, \dots , k} , B_{k+1} ] = 0.$
Therefore
\begin{eqnarray}
  \|  [ U_{t}^\dagger  S_{kt} A_{1, \dots , k} S_{kt}^\dagger U_{t} , B_{k+1} ] \|_\infty
  \le  \int_0^t \! \! \| c_{k \tau} \|_\infty  \, d\tau 
\,  \le \,  2 \| A_{1, \dots , k} \|_\infty \! \int_0^t  \|  \, [ \WFRAME{k}(\tau ) , \, B_{k+1} ] \,  \|_\infty \, d\tau  .  \ \ \ \ \ \ \ 
 \end{eqnarray}
 Separating out terms in $\WFRAME{k}$ that might become large at short times due to noncommutativity with $B_{k+1}$, and using $\| {\vec \sigma}_i \cdot {\vec \sigma}_j  \|_{\infty} = 3$, leads to
\begin{eqnarray}
 \Gamma_{\! 1t}  & \le & \  \frac{12 J_0 \, t }{n-1}  + 6 J_0 \!  \int_0^t  \!   dt_1 \, \Gamma_{ \! 2 t_{1} } 
 \label{k=1 inequality for iteration} \\
  \Gamma_{\! 2t}  & \le & \  \frac{12 (1\!+\! |\lambda|) J_0 \, t }{n-1}  + 6 (1\!+\! |\lambda|)  J_0  \! \int_0^t  \!   dt_1 \, \Gamma_{ \! 2 t_{1} } .
\label{k=2 inequality for iteration}
\end{eqnarray}

First we solve  (\ref{k=2 inequality for iteration})  iteratively, obtaining a bound for $  \Gamma_{\! 2t} $. Then we use  (\ref{k=1 inequality for iteration}) to bound $  \Gamma_{\! 1t} $. After $q$ iterations we have
\begin{eqnarray}
 \Gamma_{\! 2t} \le
 \frac{2}{ n-1 }
 \sum_{\ell=1}^q \frac{  \big( 6 (1 \! + \! |\lambda|) J_0 \,  t \big)^\ell } {\ell !}
+ \big( 6 (1 \! + \! |\lambda|) J_0 \big)^q
 \int_0^t  \! dt_{1}   \!  \int_0^{t_1}  \!  \! dt_{2}  \cdots  \int_0^{t_{q-1}} \!  \! \! dt_{q} \  \Gamma_{ \! 2 t_{q}} 
\end{eqnarray}
or
\begin{eqnarray}
 \Gamma_{\! 2t} \le
 \frac{2}{ n-1 }
 \sum_{\ell=1}^q \frac{  \big( 6 (1 \! + \! |\lambda|) J_0 \,  t \big)^\ell } {\ell !}
+ 2  \, \frac{ \big( 6 (1 \! + \! |\lambda|) J_0 t \big)^q}{q!},
\end{eqnarray}
using (\ref{lemma 1 upper bound}). 
In the large $q$ limit, 
\begin{eqnarray}
 \Gamma_{\! 2t} \, \le  \, 2 \,
 \frac{e^{6 (1 \! + \! |\lambda|) J_0 \,  t } - 1  }{ n-1} .
\label{k=2 after iteration}
\end{eqnarray}
Inserting this into (\ref{k=1 inequality for iteration}) and integrating leads to
 \begin{eqnarray}
 \Gamma_{\! 1t} \, \le  \, \frac{2}{ 1 \! + \! |\lambda|} 
 \frac{e^{6 (1 \! + \! |\lambda|) J_0 \,  t } - 1  }{ n-1} 
 \, \le \,  2 \, \frac{e^{6 (1 \! + \! |\lambda|) J_0 \,  t } - 1  }{ n-1} ,
  \label{k=1 after iteration}
\end{eqnarray}
as required. $\Box$
\end{proof}
 
\begin{lem}{}
Let $A_1$ and $B_2$ be Hermitian observables with support exclusively on qubits 1 and 2, respectively, of the $n$-qubit generalized CSM (\ref{central spin model}),  and let
\begin{eqnarray}
 \langle A_{1}\rangle 
\!  := \!  \BRA{\varphi}^{\otimes n}  U_t^\dagger A_{1}   U_t \KET{\varphi}^{\otimes n} \! \! , \ 
 \langle B_{2} \rangle 
\!  := \!  \BRA{\varphi}^{\otimes n}  U_t^\dagger B_{2}  U_t \KET{\varphi}^{\otimes n} \! \! , \ 
  \langle A_{1} B_{2} \rangle 
 :=  \BRA{\varphi}^{\otimes n}  U_t^\dagger A_{1}  B_{2}  U_t \KET{\varphi}^{\otimes n} \nonumber
\end{eqnarray}
be their expectations in the exact many-body state $U_t \KET{\varphi}^{\otimes n} \!  \! . $ 
Here $ \KET{\varphi} = \varphi_0 \KET{0} +  \varphi_1 \KET{1}$ is a pure 
single-qubit state with $ \varphi_{0,1} \in {\mathbb C}$ and $|\varphi_0|^2 +  |\varphi_1|^2  = 1$, and 
$U_t$ is the exact time-evolution operator (\ref{central spin model evolution}) of the CSM. Then
\begin{eqnarray}
C_t  := \! \! \sup_{A\neq 0,B\neq 0}  \frac{ |  \langle A_1 B_2 \rangle 
-   \langle A_1 \rangle  \langle B_2 \rangle  | }{\| A_{1} \|_\infty \, \| B_{2} \|_\infty }
 \le   \, 4 \, 
 \frac{e^{ 12 (1 \! + \! |\lambda| ) J_0  t } -1}{n-1} ,
\label{lemma 2 equation}
\end{eqnarray}
where the supremum is over the set of all bounded linear operators $A_{1}$ and  $B_{2}$ 
with support on qubits 1 and 2, respectively, such that  $ \| A_1 \|_\infty$ and $ \| B_2 \|_\infty$ are nonzero, and where $J_0$ is defined in 
(\ref{interaction bound}).
\label{lemma 2}
\end{lem}

\clearpage

\begin{proof}
The proof works by rewriting the correlation function on the left hand side of (\ref{lemma 2 equation}) in terms of commutators, and using Lemma \ref{lemma 1}. First note the equality
 \begin{eqnarray}
I^{\otimes n} = \KETBRA{\varphi}^{\otimes n} + \sum_{j=1}^n 
\KETBRA{\varphi}_1 \otimes \KETBRA{\varphi}_2 
\otimes \cdots
\otimes \KETBRA{\varphi}_{j-1} \otimes (I - \KETBRA{\varphi})_j \otimes I_{j+1}  \otimes \cdots \otimes  I_{n} ,  \nonumber
 \end{eqnarray}
 where $I$ is the two dimensional identity. Then  insert
 $I^{\otimes n}$ in  
 \begin{eqnarray}
\BRA{\varphi}^{\otimes n}  U_t^\dagger (A_{1}  \otimes B_{2})  U_t \KET{\varphi}^{\otimes n} 
= \BRA{\varphi}^{\otimes n}  (U_t^\dagger A_{1} U_t) (U_t^\dagger B_{2}  U_t ) \KET{\varphi}^{\otimes n} 
 \end{eqnarray}
 to obtain
\begin{eqnarray}
&& \BRA{\varphi}^{\otimes n}  U_t^\dagger A_{1}  B_{2}  U_t \KET{\varphi}^{\otimes n}
- \BRA{\varphi}^{\otimes n}  U_t^\dagger A_{1}   U_t \KET{\varphi}^{\otimes n}
 \BRA{\varphi}^{\otimes n}  U_t^\dagger B_{2} U_t \KET{\varphi}^{\otimes n}  \nonumber \\
 &=& \sum_{j=1}^n 
 \BRA{\varphi}^{\otimes n}
U_t^\dagger A_{1}   U_t \,
(\KETBRA{\varphi}_1 \otimes  \cdots
\otimes \KETBRA{\varphi}_{j-1} )  \otimes (I - \KETBRA{\varphi})_j  \,
 U_t^\dagger B_{2}   U_t
\KET{\varphi}^{\otimes n} 
 \nonumber \\
 &=& \sum_{j=1}^n 
 {\rm tr} \big( 
 \KETBRA{\varphi}^{\otimes n} 
U_t^\dagger A_{1}   U_t \,
(\KETBRA{\varphi}_1 \otimes  \cdots
\otimes \KETBRA{\varphi}_{j-1} )  \otimes (I - \KETBRA{\varphi})_j  \,
 U_t^\dagger B_{2}   U_t \big)
\end{eqnarray}
and
\begin{eqnarray}
&& \big| \BRA{\varphi}^{\otimes n}  U_t^\dagger A_{1}  B_{2}  U_t \KET{\varphi}^{\otimes n}
- \BRA{\varphi}^{\otimes n}  U_t^\dagger A_{1}   U_t \KET{\varphi}^{\otimes n}
 \BRA{\varphi}^{\otimes n}  U_t^\dagger B_{2} U_t \KET{\varphi}^{\otimes n} \big|  \nonumber \\
 & \le & \sum_{j=1}^n \big|  {\rm tr} \big( 
 \KETBRA{\varphi}^{\otimes n} 
U_t^\dagger A_{1}   U_t \,
(\KETBRA{\varphi}_1 \otimes  \cdots
\otimes \KETBRA{\varphi}_{j-1} )  \otimes (I - \KETBRA{\varphi})_j  \,
 U_t^\dagger B_{2}   U_t \big) \big|.  
\end{eqnarray}
Next, isolate the first two terms in the summation  and rewrite in terms of commutators, 
\begin{eqnarray}
&& \big| \BRA{\varphi}^{\otimes n}  U_t^\dagger A_{1}  B_{2}  U_t \KET{\varphi}^{\otimes n}
- \BRA{\varphi}^{\otimes n}  U_t^\dagger A_{1}   U_t \KET{\varphi}^{\otimes n}
 \BRA{\varphi}^{\otimes n}  U_t^\dagger B_{2} U_t \KET{\varphi}^{\otimes n} \big| \nonumber \\
&\le& \big| {\rm tr} \big( \KETBRA{\varphi}^{\otimes n} U_t^\dagger A_{1}  U_t  \big[ I - \KETBRA{\varphi}_1 \! ,  \, U_t^\dagger B_{2}  U_t \big]  \big) \big| \nonumber \\
 &+&  \big|  {\rm tr} \big( \KETBRA{\varphi}^{\otimes n} \big[ U_t^\dagger A_{1}  U_t  , \, I - \KETBRA{\varphi}_2 \big]  \KETBRA{\varphi}_1 U_t^\dagger B_{2}  U_t \, \big) \big| \nonumber \\
&+&  \sum_{j>2}^n \big|  {\rm tr} \big( \KETBRA{\varphi}^{\otimes n} \!\big[ U_t^\dagger A_{1}  U_t  , I - \KETBRA{\varphi}_j \big] \! \KETBRA{\varphi}_1 \otimes \cdots \otimes  \KETBRA{\varphi}_{j-1} \big[  I - \KETBRA{\varphi}_j , U_t^\dagger B_{2}  U_t  \big] \big) \big|  , \ \ \ \ \ \ \ \ 
 \end{eqnarray}
 using the property that $ I -  \KETBRA{\varphi}_i = ( I -  \KETBRA{\varphi}_i )^2 $ annihilates the initial state $\KET{\varphi}^{\otimes n}$. This leads to
\begin{eqnarray}
&& \big| \BRA{\varphi}^{\otimes n}  U_t^\dagger A_{1}  B_{2}  U_t \KET{\varphi}^{\otimes n}
- \BRA{\varphi}^{\otimes n}  U_t^\dagger A_{1}   U_t \KET{\varphi}^{\otimes n}
 \BRA{\varphi}^{\otimes n}  U_t^\dagger B_{2} U_t \KET{\varphi}^{\otimes n} \big| \nonumber \\
&\le& \|  \KETBRA{\varphi}^{\otimes n} U_t^\dagger A_{1}  U_t  \big[ I - \KETBRA{\varphi}_1 \! ,  \, U_t^\dagger B_{2}  U_t \big]  \|_1 \nonumber \\
 &+&  \|  \KETBRA{\varphi}^{\otimes n} \big[ U_t^\dagger A_{1}  U_t  , \, I - \KETBRA{\varphi}_2 \big]  \KETBRA{\varphi}_1 U_t^\dagger B_{2}  U_t  \, \|_1 \nonumber \\
&+&  \sum_{j>2}^n \|  \KETBRA{\varphi}^{\otimes n} \!\big[ U_t^\dagger A_{1}  U_t  , I - \KETBRA{\varphi}_j \big] \! \KETBRA{\varphi}_1 \otimes \cdots \otimes  \KETBRA{\varphi}_{j-1} \big[  I - \KETBRA{\varphi}_j , U_t^\dagger B_{2}  U_t  \big] \|_1   \nonumber  \\
& \le& \|  \KETBRA{\varphi}^{\otimes n}  U_t^\dagger A_{1}  U_t  \, \|_{1}  \ \|  \big[ I - \KETBRA{\varphi}_1 \! ,  \, U_t^\dagger B_{2}  U_t \big]  \|_{\infty} \nonumber \\
 &+&  \|  \KETBRA{\varphi}^{\otimes n} \big[ U_t^\dagger A_{1}  U_t  , \, I - \KETBRA{\varphi}_2 \big] \|_\infty \ \| \KETBRA{\varphi}_1 \, U_t^\dagger B_{2}  U_t  \, \|_1  \nonumber \\
&+&  \sum_{j>2}^n \|  \KETBRA{\varphi}^{\otimes n}  \big[ U_t^\dagger A_{1}  U_t  , I - \KETBRA{\varphi}_j \big] \|_1 \ \|  \KETBRA{\varphi}_1 \otimes \cdots \otimes  \KETBRA{\varphi}_{j-1} \big[  I - \KETBRA{\varphi}_j , U_t^\dagger B_{2}  U_t  \big] \|_\infty \nonumber \\
&\le& 
2 \, \|  A_{1} \|_\infty \ \|  B_{2} \|_\infty  \  \Gamma_{ \! 1t}  +  (n-2) 
\|  A_{1} \|_\infty \ \|  B_{2} \|_\infty 
\ \Gamma_{ \! 1t}^2 .
\end{eqnarray}
Here we have used the fact that both the operator and trace norms of a state (positive semidefinite matrix with unit trace) are equal to 1. Then
 \begin{eqnarray}
 && \big| \! \BRA{\varphi}^{\otimes n}  U_t^\dagger A_{1}  B_{2}  U_t \KET{\varphi}^{\otimes n}
- \BRA{\varphi}^{\otimes n}  U_t^\dagger A_{1}   U_t \KET{\varphi}^{\otimes n}
 \BRA{\varphi}^{\otimes n}  U_t^\dagger B_{2} U_t \KET{\varphi}^{\otimes n} \big|
  \nonumber \\
&& \le \|  A_{1} \|_\infty \ \|  B_{2} \|_\infty 
\  \big[ 2 \Gamma_{\! 1t}  + (n-1) \Gamma_{ \! 1t}^2 \big]
 \le 4 \,  \|  A_{1} \|_\infty \ \|  B_{2} \|_\infty 
 \frac{e^{ 12 (1 \! + \! |\lambda| ) J_0  t } -1}{n-1} .
 \end{eqnarray}
Hence, for any pair of observables $A_1$ and  $B_2$ with nonvanishing operator norms, it follows that 
$ \frac{ |  \langle A_1 B_2 \rangle 
-   \langle A_1 \rangle  \langle B_2 \rangle  | }{\| A_{1} \|_\infty \, \| B_{2} \|_\infty }
 \le   \, 4 \, 
 \frac{e^{ 12 (1 \! + \! |\lambda| ) J_0  t } -1}{n-1} ,$
leading to 
 (\ref{lemma 2 equation}) as required. $\Box$
\end{proof}
  
\noindent  Next we turn to the proof of Theorem \ref{erdos-schlein theorem}.

\begin{proof}
Let $A_1$ and $B_2$ be observables for qubits 1 and 2, respectively.
Use (\ref{linear integral equations}) and (\ref{nonlinear integral equations}) 
to obtain
\begin{eqnarray}
 \big| {\rm tr}_1 \big( A_1 X_1(t)  &-&  A_1 \rho_1(t) \big) \big| =  \bigg|  \sum_{\mu=1}^3  \int_0^t \! d\tau  \,  
J_\mu  \, {\rm tr}_1 \bigg( \!
( u_{\tau} u_{t}^\dagger  A_1 u_{t} u_{\tau}^\dagger ) 
\big[ \sigma_1^\mu , \, {\rm tr}_2 \big( ( 
X_1\! \otimes \! Y_2- \rho_{12}) 
\sigma_2^\mu  \big) \big] \bigg)  
\bigg|  \nonumber  \\
&=&  \bigg|  \sum_\mu \int_0^t \! d\tau  \,  
J_\mu  \, {\rm tr}_1  {\rm tr}_2 \bigg( \!
( u_{\tau} u_{t}^\dagger A_1 u_{t} u_{\tau}^\dagger) 
\big[ \sigma_1^\mu , \,  
( X_1\! \otimes \! Y_2- \rho_{12})
\sigma_2^\mu  \big] \bigg)  \bigg| \  \\
&=&  \bigg|  \sum_\mu \int_0^t \! d\tau  \,  
J_\mu  \, {\rm tr}_1  {\rm tr}_2 \bigg( \!
( X_1\! \otimes \! Y_2- \rho_{12})
\sigma_2^\mu 
\big[ u_{\tau} u_{t}^\dagger A_1 u_{t} u_{\tau}^\dagger , \sigma_1^\mu \big] \bigg)  \bigg| 
 \  \\
&\le&  J_0  \sum_\mu \int_0^t \! d\tau \,  \bigg| 
  {\rm tr}_1  {\rm tr}_2 \bigg( \!
( X_1\! \otimes \! Y_2- \rho_{12})
\sigma_2^\mu 
\big[ u_{\tau} u_{t}^\dagger A_1 u_{t} u_{\tau}^\dagger , \sigma_1^\mu \big] \bigg)  \bigg|  
\end{eqnarray}
and
\begin{eqnarray}
 \big| {\rm tr}_2 \big( B_2 Y_2(t)  &-& B_2 \rho_2(t) \big) \big| =  \bigg|  \sum_{\mu=1}^{3} \int_0^t \! d\tau  \,  
\frac{J_\mu}{n-1}  \, {\rm tr}_1 {\rm tr}_2 \bigg( \!
( u_{\tau} u_{t}^\dagger B_2 u_{t} u_{\tau}^\dagger) 
\big[ \sigma_2^\mu , \,   
(X_1\! \otimes \! Y_2- \rho_{12}) 
\sigma_1^\mu  \big] \bigg)  \nonumber  \\
 &+& \lambda (n-2)  \sum_\mu \int_0^t \! d\tau  \,  
\frac{J_\mu}{n-1}  \, {\rm tr}_2  {\rm tr}_3 \bigg( \!
( u_{\tau} u_{t}^\dagger B_2 u_{t} u_{\tau}^\dagger) 
\big[ \sigma_2^\mu , \,  
( Y_2\! \otimes \! Y_3 - \rho_{23}) 
\sigma_3^\mu  \big] \bigg)  \bigg|    \\
&=&  \bigg|  \sum_\mu \int_0^t \! d\tau  \,  
\frac{J_\mu}{n-1}  \, {\rm tr}_1 {\rm tr}_2 \bigg(  \!
(X_1\! \otimes \! Y_2- \rho_{12}) 
\sigma_1^\mu \big[
u_{\tau} u_{t}^\dagger B_2 u_{t} u_{\tau}^\dagger , 
 \sigma_2^\mu \big] \bigg)  \nonumber  \\
 &+& \lambda (n-2)  \sum_\mu \int_0^t \! d\tau  \,  
\frac{J_\mu}{n-1}  \, {\rm tr}_2  {\rm tr}_3 \bigg(  \!
( Y_2\! \otimes \! Y_3 - \rho_{23}) 
\sigma_3^\mu  \big[
u_{\tau} u_{t}^\dagger B_2 u_{t} u_{\tau}^\dagger , \sigma_2^\mu \big]  \bigg) \bigg| \  \\
& \le & J_0 \frac{1}{n-1}  \sum_\mu \int_0^t \! d\tau    \bigg|  {\rm tr}_1 {\rm tr}_2 \bigg(  \!
(X_1\! \otimes \! Y_2- \rho_{12}) 
\sigma_1^\mu \big[
u_{\tau} u_{t}^\dagger B_2 u_{t} u_{\tau}^\dagger , 
 \sigma_2^\mu \big] \bigg) \bigg|  \nonumber  \\
 &+& | \lambda | J_0 \, \frac{n-2}{n-1}   \sum_\mu \int_0^t \! d\tau  \bigg|
 {\rm tr}_2  {\rm tr}_3 \bigg(  \!
( Y_2\! \otimes \! Y_3 - \rho_{23}) 
\sigma_3^\mu  \big[
u_{\tau} u_{t}^\dagger B_2 u_{t} u_{\tau}^\dagger , \sigma_2^\mu \big]  \bigg) \bigg| .
\end{eqnarray}
Using the identities
\begin{eqnarray}
 X_1  \otimes  Y_2 
 &=&   (X_1- \rho_1) \otimes Y_2
 + \rho_1 \otimes  (Y_2 - \rho_2 )
 + \rho_1 \otimes  \rho_2 , \\
 Y_2  \otimes Y_3 
 &=&   (Y_2 - \rho_2) \otimes Y_3
 + \rho_2 \otimes  (Y_3 - \rho_3 )
 + \rho_2 \otimes  \rho_3 ,
\end{eqnarray}
leads to
 \begin{eqnarray}
 \big| {\rm tr} \big( A_1 (X &-& \rho_1)\big) \big|  \le  J_0  \sum_\mu \int_0^t \! d\tau \  
 \| [u_{\tau} u_{t}^\dagger A_1 u_{t} u_{\tau}^\dagger , \sigma_1^\mu]  \|_{\infty} \
 \| \sigma_2^\mu   \|_{\infty} 
 \bigg\{  \| X-\rho_{1} \|_1 \ \| Y   \|_{1}  \nonumber \\
 &+&  \| Y -\rho_{2} \|_1 \ \| \rho_1   \|_{1} 
 + \frac{| \langle  [ u_{\tau} u_{t}^\dagger A_1 u_{t} u_{\tau}^\dagger , \sigma_1^\mu] \sigma_2^\mu \rangle
  - \langle  [ u_{\tau} u_{t}^\dagger A_1 u_{t} u_{\tau}^\dagger , \sigma_1^\mu]  \rangle \langle \sigma_2^\mu  \rangle  |}{ \| [ u_{\tau} u_{t}^\dagger A_1 u_{t} u_{\tau}^\dagger , \sigma_1^\mu]  \|_{\infty} \
 \| \sigma_2^\mu   \|_{\infty} } 
 \bigg\} \\
 &\le&  6 J_0 \, \| A_1   \|_{\infty}
 \int_0^t \! d\tau \bigg\{  
 \| X-\rho_{1} \|_1  + \| Y-\rho_{2} \|_1 
 +  4 \,  \frac{ e^{12 (1 \! + \! |\lambda| ) J_0  \tau } -1}{n-1}   \bigg\} ,
 \end{eqnarray}
where $\langle \cdot \rangle = {\rm tr} (\rho \, \cdot \, )$ denotes expectation in the state 
$ \rho = U_t \big( \KETBRA{\varphi}^{\otimes n} \! \big) U_t^\dagger$. 
Similarly,
 \begin{eqnarray}
 \big| {\rm tr} \big( B_2 (Y &-& \rho_2)\big) \big|  
 \le  \frac{6 J_0 \, \| B_2  \|_{\infty} }{n-1}
 \int_0^t \! d\tau   \bigg\{  
 \| X-\rho_{1} \|_1  + \| Y-\rho_{2} \|_1 
 + 4 \,  \frac{ e^{12 (1 \! + \! |\lambda| ) J_0  \tau} -1}{n-1} 
 \nonumber  \\
&+& | \lambda |  (n-2) \bigg(  \! 2  \, \| Y-\rho_{2} \|_1
 +4 \, \frac{ e^{ 12 (1 \! + \! |\lambda| ) J_0  \tau } -1}{n-1}  \bigg) \bigg\}.
\end{eqnarray}
Assuming $\| A_1   \|_{\infty} \neq 0$ and  $\| B_2 \|_{\infty} \neq 0$,
 \begin{eqnarray}
\frac{  \big| {\rm tr} \big( A_1 (X - \rho_1)\big) \big|  }{ \| A_1   \|_{\infty} }  \le   6 J_0 \, 
 \int_0^t \! d\tau \bigg\{  
 \| X-\rho_{1} \|_1  + \| Y-\rho_{2} \|_1 
 +  4 \,  \frac{ e^{12 (1 \! + \! |\lambda| ) J_0  \tau } -1}{n-1}   \bigg\} , \\
\frac{  \big| {\rm tr} \big( B_2 (Y - \rho_2)\big) \big| }{  \| B_2  \|_{\infty} }  
 \le  \frac{6 J_0 }{n-1}
 \int_0^t \! d\tau   \bigg\{  
 \| X-\rho_{1} \|_1  + \| Y-\rho_{2} \|_1 
 + 4 \,  \frac{ e^{12 (1 \! + \! |\lambda| ) J_0  \tau} -1}{n-1} 
 \nonumber  \\
+  |\lambda| (n-2) \bigg(  \! 2  \, \| Y-\rho_{2} \|_1
 +4 \, \frac{ e^{ 12 (1 \! + \! |\lambda| ) J_0  \tau } -1}{n-1}  \bigg) \bigg\}.
 \end{eqnarray}
These hold for any $A_1$ and $B_2$ such that  $\| A_1   \|_{\infty} \neq 0$ and  $\| B_2 \|_{\infty} \neq 0$. 
Therefore
\begin{eqnarray}
\sup_{A \neq 0 }
\frac{  \big| {\rm tr} \big( A_1 (X - \rho_1)\big) \big|  }{ \| A_1   \|_{\infty} }  \le   6 J_0 \, 
 \int_0^t \! d\tau \bigg\{  
 \| X-\rho_{1} \|_1  + \| Y-\rho_{2} \|_1 
 +  4 \,  \frac{ e^{12 (1 \! + \! |\lambda| ) J_0  \tau } -1}{n-1}   \bigg\} , \\
\sup_{B \neq 0} \frac{  \big| {\rm tr} \big( B_2 (Y - \rho_2)\big) \big| }{  \| B_2  \|_{\infty} }  
 \le  \frac{6 J_0 }{n-1}
 \int_0^t \! d\tau   \bigg\{  
 \| X-\rho_{1} \|_1  + \| Y-\rho_{2} \|_1 
 + 4 \,  \frac{ e^{12 (1 \! + \! |\lambda| ) J_0  \tau} -1}{n-1} 
 \nonumber  \\
+  | \lambda |  (n-2) \bigg(  \! 2  \, \| Y-\rho_{2} \|_1
 +4 \, \frac{ e^{ 12 (1 \! + \! |\lambda| ) J_0  \tau } -1}{n-1}  \bigg) \bigg\} .
 \end{eqnarray}
Then, after using (\ref{variational trace norm equation}),
\begin{eqnarray}
\| X - \rho_1 \|_{1}  \le   6 J_0 \, 
 \int_0^t \! d\tau \bigg\{  
 \| X-\rho_{1} \|_1  + \| Y-\rho_{2} \|_1 
 +  4 \,  \frac{ e^{12 (1 \! + \! |\lambda| ) J_0  \tau } -1}{n-1}   \bigg\} , \\
\| Y - \rho_2 \|_{1} 
 \le  \frac{6 J_0 }{n-1}
 \int_0^t \! d\tau   \bigg\{  
 \| X-\rho_{1} \|_1  + \| Y-\rho_{2} \|_1 
 + 4 \,  \frac{ e^{12 (1 \! + \! |\lambda| ) J_0  \tau} -1}{n-1} 
 \nonumber  \\
+  | \lambda |  (n-2) \bigg(  \! 2  \, \| Y-\rho_{2} \|_1
 +4 \, \frac{ e^{ 12 (1 \! + \! |\lambda| ) J_0  \tau } -1}{n-1}  \bigg) \bigg\} .
 \end{eqnarray}
Up to this point in the proof we have assumed that  $n \ge 2$. If $n \gg 1$,
\begin{eqnarray}
\| X - \rho_1 \|_{1} & \le &   
 2 \, \frac{e^{ 12( 1 \! + \! |\lambda|) J_0 t } \! - \!  1}{n (1 \! + \! |\lambda|)  }
+ 6 J_0 \!
 \int_0^t \! dt_{1} \bigg(  
 \| X-\rho_{1} \|_1  + \| Y-\rho_{2} \|_1 \!  \bigg)
+ O(1/n^2) , \ \ \\
\| Y - \rho_2 \|_{1}  &\le& 
2 |\lambda| \, \frac{e^{ 12 (1 \! + \! |\lambda| ) J_0 t } \! - \!  1}{n (1 \! + \! |\lambda|) }
+  6 J_0 \!
 \int_0^t \! dt_{1}   
 \bigg(  
\frac{  \| X-\rho_{1} \|_1 }{n} +
\big( {\textstyle \frac{1}{n}} + 2 |\lambda| \big) \,
 \| Y-\rho_{2} \|_1  \! \bigg)
 +  O(1/n^2) . \ \ \ \ \ \ 
 \end{eqnarray}
 We solve these iteratively. After $q$ iterations we have
\begin{eqnarray}
 \| X - \rho_1 \|_{1}   &\le&  
 2 \, \frac{ e^{12 (1 \! + \! |\lambda| ) J_0  t}-1 }{n (1 \! + \! |\lambda|)}  
 \bigg[ 1 + (a_1 \! + \! |\lambda| \,  b_1)  \times \!  
 \bigg( \! \frac{1}{ 2 (1 \! + \! |\lambda|) } \! \bigg)
 + \cdots +  (a_{q-1} \! + \!  |\lambda| \, b_{q-1} ) 
 \bigg( \! \frac{1}{ 2 (1 \! + \! |\lambda|) } \! \bigg)^{\! q-1}
\bigg] \nonumber \\
&+& \, (6 J_0)^q  \! \! 
\int_0^{t}  \! \! dt_1  
\cdots  \!  \int_0^{t_{q-1}}  \! \! \! dt_q
\bigg[ a_q \,  \| X - \rho_1 \|_{1}  + 
 b_q \,  \| Y - \rho_2 \|_{1} 
\bigg] +  O(1/n^2),
\label{after q iterations x}
\end{eqnarray}
and 
\begin{eqnarray}
 \| Y - \rho_2 \|_{1}   &\le&  
 2 \, \frac{ e^{12 (1 \! + \! |\lambda|) J_0  t}-1 }{n (1 \! + \! |\lambda|)}  
 \bigg[ |\lambda| + ( a^{\prime}_1 + |\lambda| \,  b^{\prime}_1) 
 \bigg( \! \frac{1}{ 2 (1 \! + \! |\lambda|) } \! \bigg)
 + \cdots +  (a^{\prime}_{q-1} +  |\lambda|  \, b^{\prime}_{q-1} ) 
 \bigg( \! \frac{1}{ 2 (1 \! + \! |\lambda| )  } \! \bigg)^{\! q-1}
\bigg] \nonumber \\
&+& \, (6 J_0)^q  \! \! 
\int_0^{t}  \! \! dt_1  
\cdots  \!  \int_0^{t_{q-1}}  \! \! \! dt_q
\bigg[ a^{\prime}_q \,  \| X - \rho_1 \|_{1}  + 
 b^{\prime}_q \,  \| Y - \rho_2 \|_{1} 
\bigg] +  O(1/n^2),
\label{after q iterations y}
\end{eqnarray}
where the positive real coefficients $a_k, b_k$ satisfy
\begin{eqnarray}
a_1= 1, \ \ b_1 = 1,
\label{recurrence relation initial condition 1} 
\end{eqnarray}
and
\begin{eqnarray}
a_k = a_{k-1} + \frac{b_{k-1}}{n} ,
\label{recurrence relation a} \\
b_k = a_{k-1} + m \, b_{k-1}  ,
\label{recurrence relation b}
\end{eqnarray}
for $k>1$, where
\begin{eqnarray}
m := \frac{1}{n} + 2 | \lambda | .
\label{m definition}
\end{eqnarray}
The coefficients $a^{\prime}_k , b^{\prime}_k$ in (\ref{after q iterations y}) satisfy the identical recurrence relation 
 but start with
\begin{eqnarray}
a^{\prime}_1= \frac{1}{n}, \ \ 
b^{\prime}_1 = m,
\label{recurrence relation initial condition 2} 
\end{eqnarray}
instead of (\ref{recurrence relation initial condition 1}). Equations (\ref{recurrence relation a}) and (\ref{recurrence relation b}) can be solved for 
arbitrary $a_1, b_1$:
\begin{eqnarray}
a_k &=& \bigg[ 1 + \frac{1 + (1+m) + (1 + m + m^2) + \cdots 
+ (1 + m + m^2 +m^3 + \cdots m^{k-3} )  }{n} 
\bigg] a_1  \nonumber \\
 &+& \bigg[  \frac{1 + m + m^2 +m^3 + \cdots m^{k-2}  }{n} 
\bigg] b_1
+  O(1/n^2) \\
&=& \bigg[  1 + \frac{ 1 - 2m + (k-3)(1-m) + m^{k-1} }{  n (1-m)^2 } \bigg] a_1 +
 \frac{ 1 - m^{k-1} }{n (1-m)  } \, b_1
+  O(1/n^2) ,
\end{eqnarray}
\begin{eqnarray}
b_k &=& \bigg[ \frac{1-m^{k-1} }{1-m} + 
\frac{  (k-3) \, m^{k+1} + (1-k) \, m^k + 2 m^3 - m^2 + (k-1) \, m + 2 - k } {  n m^2 (m-1)^3 }
\bigg] a_1  \nonumber \\
 &+& \bigg[  m^{k-1} + 
 \frac{  1 - m^{k-1} + (k-1)(m-1) m^{k-2} } 
{ n (1-m)^2 } 
\bigg] b_1
+  O(1/n^2) .
\end{eqnarray}
Anticipating the large $n$ limit, we have dropped terms $1/n^2$ and smaller. The second forms of the above expressions are obtained by assuming $m \neq 1$ and summing geometric series and their derivatives. Note that for $ (a_1,  b_1) = (1,1)$, we have
\begin{eqnarray}
a_k + |\lambda| \, b_k = 1 + |\lambda|  \frac{1 - m^k}{1-m} + O(1/n) ,
\label{combination 1}
\end{eqnarray}
whereas for $ (a'_1,  b'_1) = (\frac{1}{n},m)$ we have
\begin{eqnarray}
a'_k + |\lambda| \, b'_k = |\lambda| \, m^k + O(1/n) .
\label{combination 2}
\end{eqnarray}
Using (\ref{combination 1}) and (\ref{combination 2}),
\begin{eqnarray}
&& \lim_{n \rightarrow \infty}  \sum_{k=1}^{q-1} 
\frac{a_k + |\lambda| \, b_k}{ (2+ 2|\lambda|)^k } 
=  \frac{1}{1-2 | \lambda| } \sum_{k=1}^{q-1} 
\frac{  (1 -  | \lambda |)  - |\lambda|  (2 | \lambda|)^k  }{ (2+ 2|\lambda|)^k   } + O(1/n) ,  \\
&& \lim_{n \rightarrow \infty}  \sum_{k=1}^{q-1} 
\frac{a'_k + |\lambda| \, b'_k}{ (2+ 2|\lambda|)^k } 
=  |\lambda|  \sum_{k=1}^{q-1} 
\frac{ |2 \lambda|^k  }{ (2+ 2|\lambda|)^k   } + O(1/n) .
\end{eqnarray}
Then we obtain, for $ | \lambda | \le 1$,
\begin{eqnarray}
\lim_{q \rightarrow \infty} 
\lim_{n \rightarrow \infty} 
\bigg( 1 + \sum_{k=1}^{q-1} 
\frac{a_k + |\lambda| \, b_k}{ (2+ 2 |\lambda| )^k } \bigg)  
\ \le 1 + \frac{  1 -   |\lambda|  -  |\lambda|^2 - 2 | \lambda|^3   } { (1 + 2 |\lambda| ) (1 - 2 |\lambda| ) }
\le 2
\end{eqnarray}
and
\begin{eqnarray}
\lim_{q \rightarrow \infty} 
\lim_{n \rightarrow \infty} 
\bigg(  |\lambda|  + \sum_{k=1}^{q-1} 
\frac{a'_k + |\lambda| \, b'_k}{ (2+ 2 |\lambda| )^k } \bigg)  \le
|\lambda|  +  \lambda^2
\le 2.
\end{eqnarray}
Finally, note that 
\begin{eqnarray}
 (6 J_0)^q  \! \! 
\int_0^{t}  \! \! dt_1  
\cdots  \!  \int_0^{t_{q-1}}  \! \! \! dt_q
\bigg[ a_q \,  \| X - \rho_1 \|_{1}  + 
 b_q \,  \| Y - \rho_2 \|_{1} 
\bigg] \le 
2  \,   (a_q + b_q)   \,  \frac{ (6 J_0 t )^q}{q!}     \\
 (6 J_0)^q   \! \! 
\int_0^{t}  \! \! dt_1  
\cdots  \!  \int_0^{t_{q-1}}  \! \! \! dt_q
\bigg[ a^{\prime}_q \,  \| X - \rho_1 \|_{1}  + 
 b^{\prime}_q \,  \| Y - \rho_2 \|_{1} 
\bigg]  \le
2 \,  (a'_q + b'_q)   \,  \frac{ (6 J_0 t )^q}{q!} 
\end{eqnarray}
both vanish in the large $q$ limit. Then we obtain (\ref{erdos-schlein thm x equation}) as required. $\Box$
\end{proof}

\clearpage

\section{Discussion}
\label{discussion section}

Mean field errors are bounded by a competition between an exponential growth in time and a $1/n$ suppression in system size, so the bounds are mainly interesting when $n \gg {\rm exp}(O(t))$. 
Thus, it is tempting to conclude that the CSM requires exponentially many qubits to simulate nonlinearity, but this is not the case for a finite-time simulation. This can be understood by assuming $ 12 (1 \! + \! |\lambda| ) J_0  t  \ll 1$, which defines a particular short-time limit, and linearizing the exponential in (\ref{erdos-schlein thm x equation}). This leads to
\begin{eqnarray}
\| X(t) - \rho_1  (t) \|_1
\le
 \frac{ 48 J_0  t }{ n } 
 = \epsilon ,
\end{eqnarray}
where $\epsilon$ is the desired {\it model} error. 
Then duality within $\epsilon$ holds for a time 
\begin{eqnarray}
t_{\rm max} = \frac{n \epsilon}{48 J_0}  = 
n\, \Delta t , \ \ \Delta t := \frac{\epsilon}{48 J_0} .
\end{eqnarray}
In the short-time regime, increasing $n$  merely increases the simulation interval $ t_{\rm max} $, each ancilla qubit contributing a unit of propagation time $\Delta t$.

If $\lambda = 1$ and complete permutation symmetry is respected,  the CSM is described 
by mean field theory (\ref{self interaction model}), which has self-interaction. This nonlinearity generates qubit torsion and other nonrigid distortions of the Bloch ball determined by the couplings $J_\mu$ \cite{190709349,211113477}. 
To see this, write the Hamiltonian in (\ref{self interaction model}) as
\begin{eqnarray}
H^{\rm eff}  
 = H^0 + \sum_\mu J_\mu \, {\rm tr}( X  \sigma^\mu) \, \sigma^\mu,
\label{self interaction hamiltonian discussion}
\end{eqnarray}
where $X$ is the current state of the central (or any other)  qubit.
Suppose $J_\mu = (J_1, 0, 0)$. The nonlinear term in (\ref{self interaction hamiltonian discussion}) generates an $x$ rotation with frequency $2 J_1 x$, where  $x$ is the projection of the Bloch vector on the $x$ axis. States with larger $x$ components rotate faster, and states with negative projections rotate in the opposite direction, twisting the Bloch ball. Couplings $(0, J_2, 0)$ and $(0, 0, J_3)$ similarly generate pure torsion about the $y$ and $z$ axes of the Bloch ball, respectively. Single-axis torsions have been investigated previously \cite{PhysRevLett.81.3992,150706334,211105977}. More general couplings $J_\mu = (J_1, J_2, J_3)$ with two or three nonzero components generate higher-order distortions 
beyond pure torsion, which have not been studied.


\clearpage

The CSM with $\lambda \neq 1$  is described by the coupled nonlinear equations 
 (\ref{x evolution equation}) and
  (\ref{y evolution equation}).  
 The CSM with $\lambda = 0$ is particularly interesting: In this case the Hamiltonian for the central qubit is
\begin{eqnarray}
H^{\rm eff}  
 = H^0 + \sum_\mu J_\mu \, {\rm tr}( Y  \sigma^\mu) \, \sigma^\mu \! ,
\label{central hamiltonian discussion}
\end{eqnarray}
where, in the large $n$ limit, $Y$ is governed by $H^0$ only. Thus, the central qubit interacts with a bath of synchronized ancilla, but produces vanishing reaction on any individual ancilla qubit. To use this for information processing, set $H^0 =0$. Then $\frac{dY}{dt}=0$ and the resulting Hamiltonian 
\begin{eqnarray}
H^{\rm eff}  
 = \sum_\mu J_\mu \BRA{\varphi}  \sigma^\mu \KET{\varphi}   \sigma^\mu 
\label{central hamiltonian interaction discussion}
\end{eqnarray}
implements 
initial-condition nonlinearity 
($ \langle \sigma^\mu \rangle$
 is static and fixed by the initial condition). Different initial states $\KET{\varphi}$ 
are subjected to 
different Hamiltonians. If $J_\mu $ is time-independent, these are static Hamiltonians, whereas (\ref{self interaction hamiltonian discussion})  is typically time dependent (because $X$ is).

Finally, we speculate on the relevance of the duality to the question of whether quantum mechanics is fundamentally nonlinear. While there is no experimental evidence for such nonlinearity 
\cite{WeinbergAP89,WeinbergPRL89,BollingerPRL89,PhysRevLett.64.2261,WalsworthPRL90,MajumderPRL90,190901608}, it would be more illuminating to have a theoretical argument or no-go theorem showing that its presence would violate a stronger property, such as relativistic invariance \cite{GisinPLA90,PolchinskiPRL91,GisinJPA95,CzachorPLA02,KentPRA05}. However no such argument is currently available. Dualities like that discussed here suggest that there might not be a sharp distinction between universes evolving according to linear and nonlinear quantum mechanics. This observation is consistent both with the absence of a nonlinear no-go theorem and with other dualities based on nonlinear
gauge transformations \cite{DoebnerJMP99}.
If quantum nonlinearity is indeed allowed, how can we experimentally test for it? Beyond laboratory experiments \cite{WeinbergAP89,WeinbergPRL89,BollingerPRL89,PhysRevLett.64.2261,WalsworthPRL90,MajumderPRL90,190901608}, one possibility  is to consider the cosmological implications of potential quantum nonlinearity \cite{KibbleCMP78,9512004,13124455,200808663,210610576}. Lloyd \cite{13124455} has argued that the universe itself might be regarded as a giant quantum information processor, and that this perspective explains how the complexity observed today could arise from a homogeneous, isotropic initial state evolving according to ``simple" laws. In the future it would be interesting to reexamine the question of cosmological complexity generation with the hypothesis of real or simulated quantum nonlinearity.

\clearpage

\clearpage

\acknowledgements
This work was partly supported by the NSF under grant no.~DGE-2152159. It is a pleasure to thank Benjamin Schlein for correspondence.

\appendix

\section{Partial traces of commutators}

\noindent Here we explain some  properties of partial traces used in the proofs.

\begin{enumerate}

\item
Let $\rho \in \BLO$ be any bounded linear operator, and let $B_i$ be an operator acting on qubit $i$ exclusively. Then the partial trace of their commutator vanishes:
\begin{eqnarray}
{\rm tr}_{i} (  [B_i, \rho] ) = 0.
\label{first partial trace identity} 
\end{eqnarray}
To see this, evaluate ${\rm tr}_{i}([B_i, \rho]) $ in the $\{  \KET{0} , \KET{1} \}$ basis of qubit $i$:
\begin{eqnarray}
{\rm tr}_{i}([B_i, \rho]) 
&=& \sum_{x,x'=0,1} 
\! \bigg( \! \!
\BRA{x} B_i \KET{x'}_i
\BRA{x'} \rho \KET{x}_i
- \BRA{x} \rho \KET{x'}_i
\BRA{x'} B_i \KET{x}_i  \! \bigg) \\
&=& \sum_{x,x'=0,1} 
\! \bigg( \! \!
\BRA{x} B_i \KET{x'}_i
\BRA{x'} \rho \KET{x}_i
- \BRA{x'} \rho \KET{x}_i
\BRA{x} B_i \KET{x'}_i  \! \bigg) \\
&=& \sum_{x,x'=0,1} 
\BRA{x} B_i \KET{x'}_i
\! \bigg( \! \!
\BRA{x'} \rho \KET{x}_i
- \BRA{x'} \rho \KET{x}_i  \! \bigg) 
= 0,
 \end{eqnarray}
because $\BRA{x} B_i \KET{x'}_i \in {\mathbb C}$ commutes with the operator $\BRA{x'} \rho \KET{x}_i. $

\item 
Let $\rho \in \BLO$ be any bounded linear operator, and let $B_i$ be an operator acting on qubit $i$ exclusively. Then
\begin{eqnarray}
{\rm tr}_{>j}([B_i , \rho])  = 
{\rm tr}_{j+1} {\rm tr}_{j+2}  \cdots {\rm tr}_{n}([B_i , \rho]) =
\bigg\lbrace
\begin{array}{lr}
[B_i ,  {\rm tr}_{>j}(\rho) ] , & \text{for } i \leq j \\
0, & \text{for } i > j 
\end{array} .
\label{second partial trace identity}
\end{eqnarray}
If $i \le j $ then ${\rm tr}_{j+1} \cdots {\rm tr}_{n}( B_i  \rho -  \rho B_i  ) = [B_i ,  {\rm tr}_{>j}(\rho) ]. $ 
If $ i > j $ the required result follows from 
(\ref{first partial trace identity}). 
 
\end{enumerate}

\clearpage

\section{Schatten $p$-norms}

Here we collect a few properties of the matrix norms used in this paper.  Let $X \in {\mathbb C}^{2^n \times 2^n}$ be a complex matrix on $n$ qubits. The norms $\| X \|_{1}$  and $\| X \|_{\infty}$  used in Theorem \ref{erdos-schlein theorem}  
(Sec.~\ref{large n limit section}) are special cases of Schatten $p$-norms
\begin{eqnarray}
\| X \|_p := [{\rm tr}(|X|^p)]^\frac{1}{p} , \ \ 
p \ge 1,
\end{eqnarray}
where $|X| := \sqrt{X^\dagger X}$
is the absolute value of a matrix. 
Because $A = X^\dagger X = U D U^\dagger$ is Hermitian and positive semidefinite, we can define $\sqrt{A} = U \sqrt{D} U^\dagger$ through its spectral decomposition, leading to
$|X|  = U \sqrt{D} \, U^\dagger = U \Sigma U^\dagger, $
where $\Sigma$ is a diagonal matrix containing the singular values 
$  \sqrt{ {\rm spec}(X^\dagger X)} $
of $X$. 
Here ${\rm spec}(Y)$ denotes the set of eigenvalues of $Y \in \BLO$, and $\sqrt{{\rm spec}(Y)}$ are their square roots. Then
$ \| X \|_p  = [  {\rm tr}(\Sigma^p) ]^\frac{1}{p} 
=  [ \sum_{i=1}^{2^n}  ( \Sigma_{ii})^p ]^{\! \frac{1}{p} } $. 

\vspace{0.2in}

\noindent We use the following properties:

\begin{enumerate}

\item 
The Schatten $p$-norm is unitarily invariant. Let $U,V \in {\mathbb C}^{ 2^n \times 2^n }$ be unitary. Then $ \| UXV^\dagger \|_p = \| X \|_p$. 

\item
The Schatten $p$-norm is submultiplicative: 
\begin{eqnarray}
 \| XY \|_p \le  \| X \|_p  \| Y \|_p .
 \label{submultiplicative inequality}
\end{eqnarray}

\item
The Schatten 1-norm  $ \| X \|_1$ is equal to the trace norm (sum of singular values).

\item
The Schatten 1-norm satisfies 
\begin{eqnarray}
  | {\rm tr}(X)| \le \| X \|_1 .
 \label{trace norm trace inequality}
\end{eqnarray}

\item
The Schatten 1-norm is not normalized:  $ \| I^{\otimes n} \|_1= 2^n$. Here $I$ is the 2-dimensional identity. 

\item
The limit
$ \| X \|_\infty := \lim_{p \rightarrow \infty}  \| X \|_p$ exists and is equal to the operator norm (maximum singular value).

\item
The operator norm is normalized:  $ \| I^{\otimes n} \|_\infty =  \| I \|_\infty  = 1$. 

\item
The trace and operator norms satisfy the inequality 
\begin{eqnarray}
 \| X \|_\infty  \le   \| X \|_1 .
 \label{trace and operator norm inequality}
\end{eqnarray}

\item
The trace and operator norms also satisfy a Holder inequality 
\begin{eqnarray}
  \| XY \|_1 \le  \| X \|_1  \| Y \|_\infty ,
 \label{holder inequality}
\end{eqnarray}
which is tighter than that provided by (\ref{submultiplicative inequality}). 

\item
Let $A \in \BLO$ be a bounded linear operator. Then
\begin{eqnarray}
 \sup_{B \neq 0}
\frac{  | {\rm tr}( AB) |  }{ \| B \|_\infty } = \| A \|_1 ,
\label{variational trace norm equation}
\end{eqnarray}
where the supremum is over the set of all $ B \in \BLO $ with $ \| B \|_\infty \neq 0$. 

\item
Let $X_\alpha , X_\beta$ be arbitrary states (positive semidefinite operators with unit trace). Then
\begin{eqnarray}
\| X_\alpha - X_\beta  \|_1 \le 2.
\label{trace distance bound}
\end{eqnarray}

\item 
Let $A, B \in \CMATRIX{\DIM}$ and  $C \in \CMATRIX{\DIM^2} \!  .$ 
Then
\begin{eqnarray}
\int_0^t \! d\tau \,  \big| {\rm tr} \big( C \cdot 
A \! \otimes \!  B \big) \big|  
\le  \int_0^t \! d\tau \,  \| C(\tau) \|_1 \, \| A(\tau) \|_{\infty} \, \| B(\tau) \|_{\infty} , \\
\int_0^t \! d\tau \,  \big| {\rm tr} \big( C \cdot 
A \! \otimes \!  B \big) \big|  
\le  \int_0^t \! d\tau \,  \| C(\tau) \|_{\infty} \, \| A(\tau) \|_{1} \, \| B(\tau) \|_{1} .
\label{integral of trace inequality}
\end{eqnarray}

\item
Let ${\vec \sigma}_i \cdot {\vec \sigma}_j = 
\sigma_{i}^{1} \otimes \sigma_{j}^1 
+ \sigma_{i}^{2} \otimes \sigma_{j}^2 
+ \sigma_{i}^{3} \otimes \sigma_{j}^3 $.
Then
\begin{eqnarray}
 \| {\vec \sigma}_i \cdot {\vec \sigma}_j  \|_{\infty} = 3 \ \ {\rm and} \ \ 
  \| {\vec \sigma}_i \cdot {\vec \sigma}_j  \|_{1} = 6.
 \label{sigma squared norms}
\end{eqnarray}

\end{enumerate}

\clearpage
 
\section{Simulations}

\begin{figure}
\includegraphics[width=16.0cm]{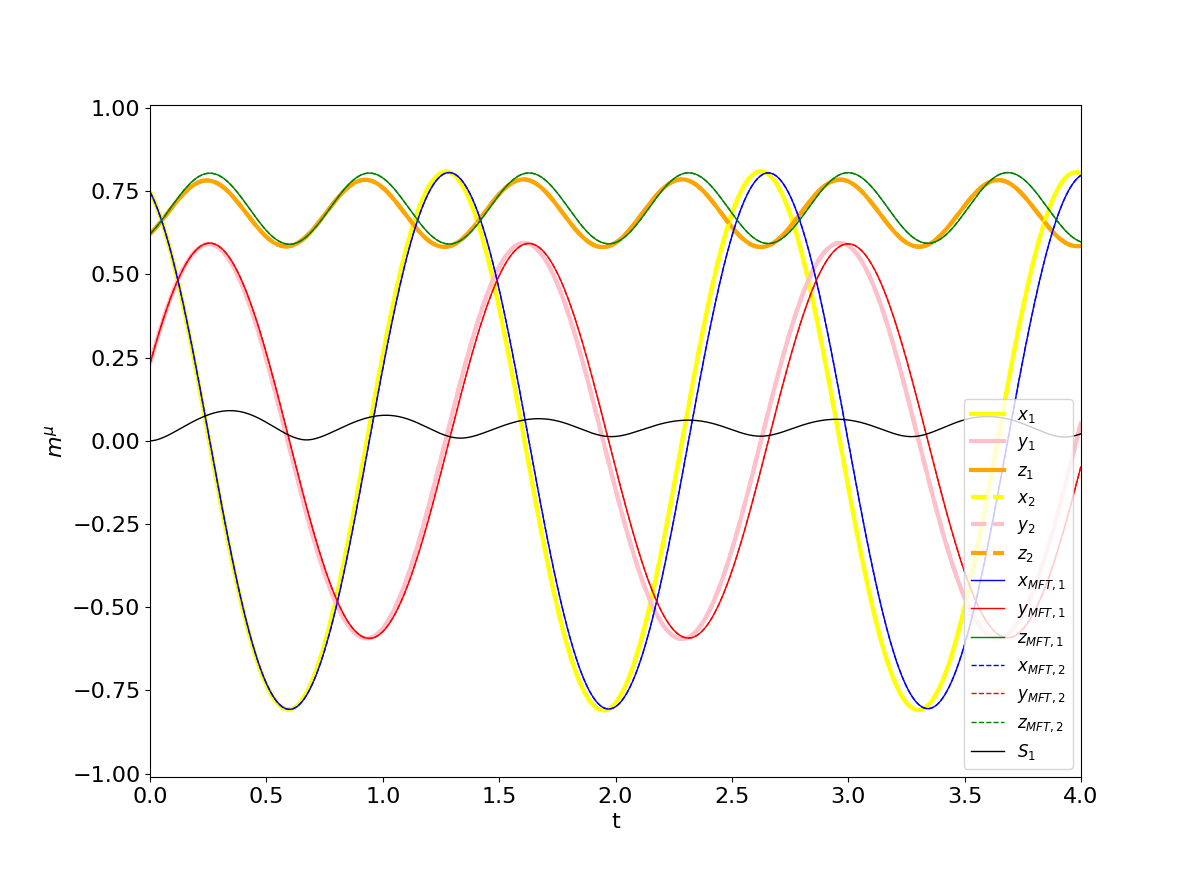} 
\caption{Bloch vector components for model (\ref{simulated model parameters}) with $\lambda = 1$ and initial condition (\ref{simulated initial condition}).}
\label{bloch lambda1 figure}
\end{figure} 

\begin{figure}
\includegraphics[width=16.0cm]{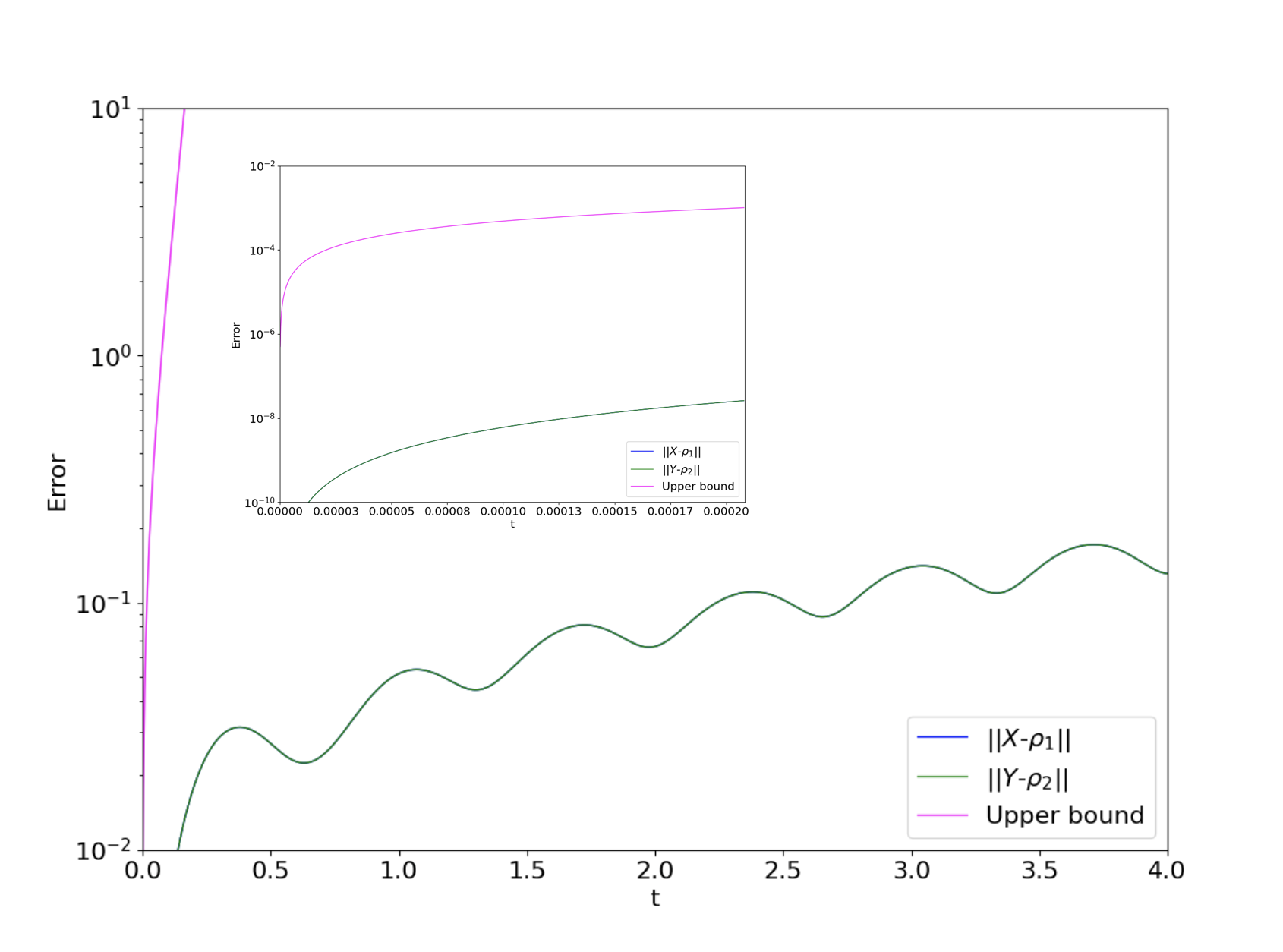} 
\caption{State errors for model (\ref{simulated model parameters}) with $\lambda = 1$ and initial condition (\ref{simulated initial condition}).}
\label{error lambda1 figure}
\end{figure}

\begin{figure}
\includegraphics[width=16.0cm]{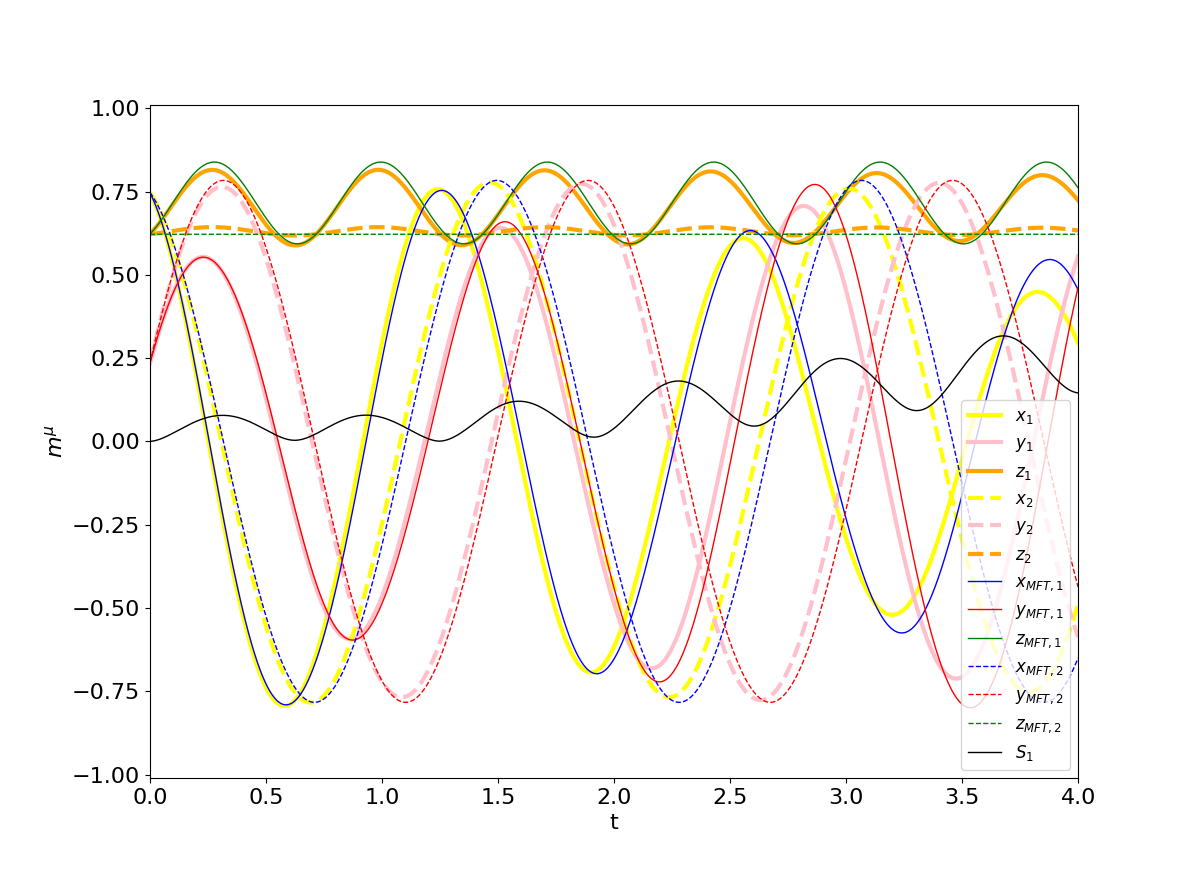} 
\caption{Bloch vector components for model (\ref{simulated model parameters}) with $\lambda = 0$ and initial condition (\ref{simulated initial condition}).}
\label{bloch lambda0 figure}
\end{figure}

\begin{figure}
\includegraphics[width=16.0cm]{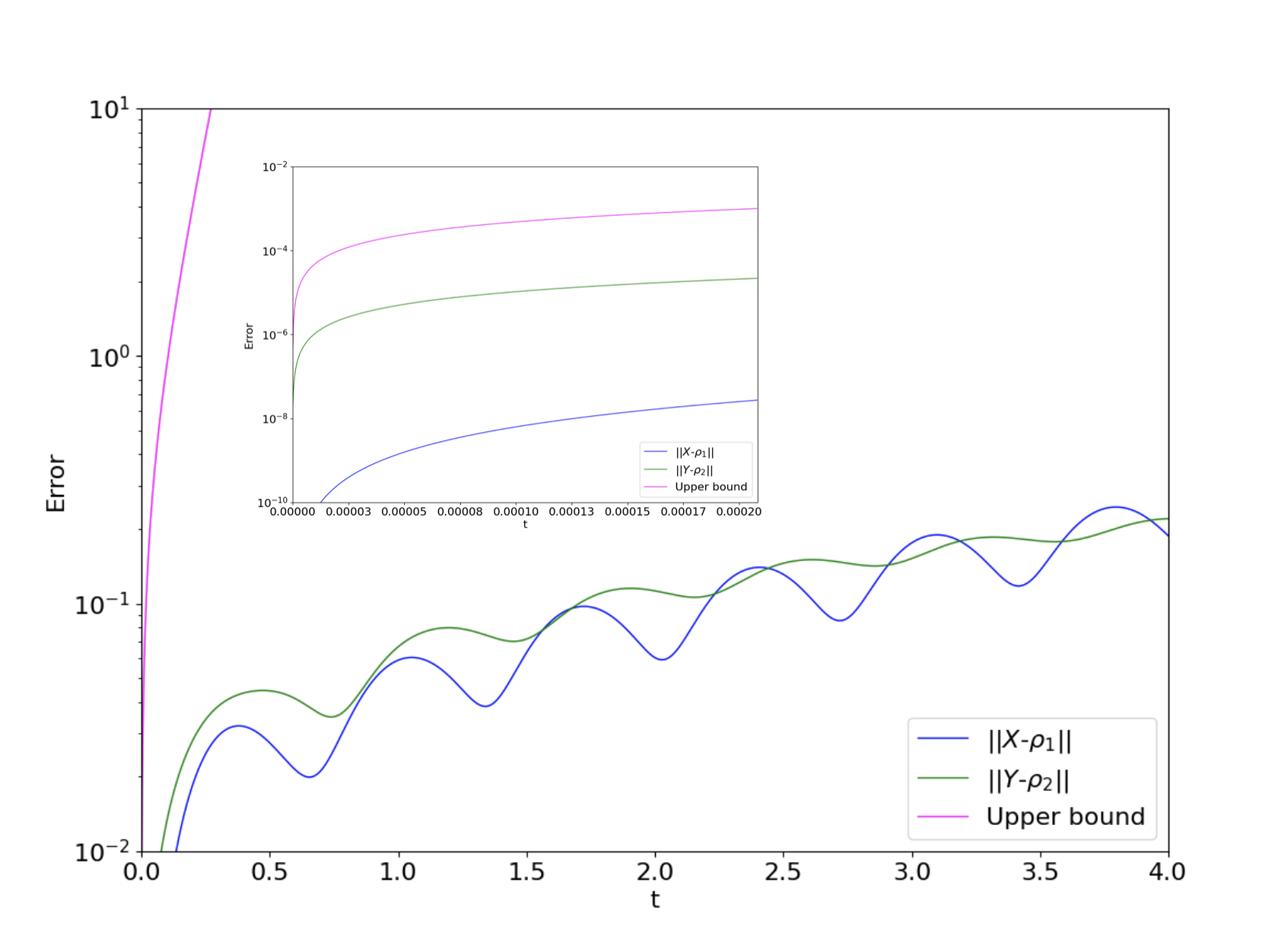} 
\caption{State errors for model (\ref{simulated model parameters}) with $\lambda = 0$ and initial condition (\ref{simulated initial condition}).}\label{error lambda0 figure}
\end{figure}

Here we show small-$n$ simulation results for two cases of the CSM, one with $\lambda = 1$ and $S_n$ symmetry (Figs.~\ref{bloch lambda1 figure}-\ref{error lambda1 figure}), the other with $\lambda = 0$ and $S_{n-1}$ symmetry (Figs.~\ref{bloch lambda0 figure}-\ref{error lambda0 figure}). Apart from these permutation symmetry assumptions, we consider a ``typical'' low-symmetry instance of the model 
\begin{eqnarray}
  \lambda = 0,1, \ \ J_\mu = (1, -1,\textstyle{\frac{1}{2}} ), \ \ J_0 = 1, \ \ B=2, \ \ n=10, 
  \label{simulated model parameters}
\end{eqnarray}
and a low-symmetry initial condition,
\begin{eqnarray}
\KET{\varphi} = \varphi_0 \KET{0} +  \varphi_1 \KET{1}, \ 
\varphi_0 = \cos(\theta/2), \ \ \varphi_1 = e^{\I \phi} \sin(\theta/2), \ \ 
\theta = 0.90, \ \ \phi=0.30.
 \label{simulated initial condition}
\end{eqnarray}

\clearpage

\leftline{}
\vspace{0.3in}

First consider the $\lambda=1$ simulation results shown in Figs.~\ref{bloch lambda1 figure}
and \ref{error lambda1 figure}. 
Here qubit 1 is the central qubit and qubit 2 is an ancilla qubit.
To read Fig.~\ref{bloch lambda1 figure},  note that the exact Bloch vector components ($x_1, \cdots , z_2$) are thicker lines, with qubit 1 solid and qubit 2 dashed. However the qubit 1 (solid) and qubit 2 (dashed) curves in this figure are identical due to permutation symmetry (so the dashed curves are not visible). Overall, mean field theory is very accurate for this 10-qubit system. The entanglement entropy (black curve) shows very little entanglement developing between the central qubit and remaining 9 ancilla. The mean field theory state errors are shown in Fig.~\ref{error lambda1 figure}. 
Upper bound is the bound (\ref{erdos-schlein thm x equation}-\ref{erdos-schlein thm x equation}).
The inset magnifies the short-time regime corresponding to model error $\epsilon = 10^{-3}$. This is the set of times where the {\it bound} is below $\epsilon$, the regime where the CSM reliably simulates  nonlinear quantum mechanics to error $\epsilon$.

\clearpage

\leftline{}
\vspace{0.3in}

Figures~\ref{bloch lambda0 figure} and \ref{error lambda0 figure} repeat this analysis for the $\lambda = 0$ CSM. The main difference is that now the central qubit and ancilla have different dynamics.  Also, the ancilla errors are usually larger than the central qubit state errors. This is a finite-size effect resulting from the $O(1/n)$ term neglected in passing from (\ref{y evolution equation})  to (\ref{y evolution equation large n}), which imparts an error on the equation of motion for the ancilla qubit $Y$, but not on the central qubit $X$. This asymmetry is especially apparent in the short-time regime.

\clearpage

\bibliography{Paper.bbl}

\end{document}